\documentclass[11pt]{article}
\usepackage{jheppub}

\usepackage{color}
\usepackage{amsmath}

\usepackage{verbatim}   
\usepackage{subfigure}  
\usepackage{amsfonts}
\usepackage{amssymb}
\usepackage{mathrsfs}
\usepackage{graphicx}
\usepackage{slashed}
\usepackage{amsthm}

\usepackage{graphicx}
\usepackage{dcolumn}
\usepackage{bm}

\usepackage{color}
\usepackage{amsmath}

\usepackage{verbatim}   
\usepackage{subfigure}  
\usepackage{amsfonts}
\usepackage{amssymb}
\usepackage{mathrsfs}
\usepackage{graphicx}
\usepackage{slashed}
\usepackage{amsthm}
\def\beq{\begin{equation}\begin{aligned}}
\def\eeq{\end{aligned}\end{equation}}

\def\d{\textrm{d}}
\def\d{\textrm{d}}
\def\tr{\textrm{tr}}

\def\a{\alpha'}
\def\A{{\cal A}}

\def\bar#1{\overline{#1}}

\def\inv{^{\raise.15ex\hbox{${\scriptscriptstyle -}$}\kern-.05em 1}}
\def\lbar{{\lower.35ex\hbox{$\mathchar'26$}\mkern-10mu\lambda}} 

\def\to{\rightarrow}

\def\p{\partial}
\def\bp{\bar\partial}

\def\R{\mathcal{R}}
\def\End{{\rm End}}

\let\p=\partial
\let\<=\langle
\let\>=\rangle

\let\+=\uparrow

\definecolor{mygreen}{RGB}{29,145,47}
\definecolor{mypurple}{RGB}{164,64,214}
\definecolor{myorange}{RGB}{199,146,32}

\newtheorem*{Theorem*}{Theorem}

\definecolor{orange}{rgb}{1,0.5,0}

\begin{document}
\title{$T^3$-Invariant Heterotic Hull-Strominger Solutions}

\author[a,b]{Bobby Samir Acharya,}
\author[c]{Alex Kinsella,}
\author[d]{and Eirik Eik Svanes}

\affiliation[a]{Department of Physics, Kings College London, London, WC2R 2LS, UK}
\affiliation[b]{Abdus Salam International Centre for Theoretical Physics, Strada Costiera 11, 34151, Trieste, Italy}
\affiliation[c]{Department of Physics, Broida Hall, University of California Santa Barbara, CA 93106}
\affiliation[d]{Department of Mathematics and Physics, University of Stavanger, Kristine Bonnevies vei 22,\\  4021 Stavanger}

\begin{flushright}
KCL-PH-TH/2020-60
\end{flushright}


\abstract{We consider the heterotic string on Calabi-Yau manifolds admitting a Strominger-Yau-Zaslow fibration. Upon reducing the system in the $T^3$-directions, the Hermitian Yang-Mills conditions can then be reinterpreted as a complex flat connection on $\mathbb{R}^3$ satisfying a certain co-closure condition. We give a number of abelian and non-abelian examples, and also compute the back-reaction on the geometry through the non-trivial $\a$-corrected heterotic Bianchi identity, which includes an important correction to the equations for the complex flat connection. These are all new local solutions to the Hull-Strominger system on $T^3\times\mathbb{R}^3$. We also propose a method for computing the spectrum of certain non-abelian models, in close analogy with the Morse-Witten complex of the abelian models.}

\maketitle

\section{Introduction}
Since the early days of string theory, heterotic compactifications have been a fruitful road towards realistic models of particle physics, beginning with the seminal paper of Candelas, Horowitz, Strominger and Witten \cite{Candelas:1985en}. It is common practice to look for supersymmetric solutions by compactifying on a six-dimensional Calabi-Yau manifold with a gauge bundle, at least as a zeroth-order geometry. However, $\a$-corrections generically induce torsion \cite{Strominger1986,Hull1986357}, whereby the geometry is described by a more complicated set of equations known as the Hull-Strominger system. The geometric features of the torsional Hull-Strominger system are much more mysterious than their Calabi-Yau cousins. Part of the purpose of this paper is to shed light on the local geometry of these solutions by considering a certain dimensional reduction of the geometry and studying local solutions to the system in this limit. In fact, the reduced solutions we present constitute new $T^3$-invariant local solutions of the Hull-Strominger system. 

Another route to physically realistic effective theories is via M-theory on singular $G_{2}$ holonomy spaces \cite{Acharya:2001gy}. Specifically, $G_{2}$ holonomy spaces with codimension-7 singularities sitting in codimension-4 orbifold loci may give a geometry on which M-theory can produce realistic models of particle physics. One tool to investigate M-theory compactifications is the duality between M-theory on a K3 surface and the $E_{8}\times E_{8}$ heterotic string on $T^{3}$ \cite{Witten:1995ex}. This duality is simplest in the limit of large heterotic volume, which corresponds on the M-theory side to the half-K3 limit, where the K3 surface is stretched along one direction (analogous to the stable degeneration limit of F-theory). This duality may be adiabatically fibered over a 3D base to obtain 4D effective theories: when the $G_{2}$ holonomy space of the M-theory geometry carries a coassociative K3 fibration, we expect it to be dual to the $E_{8}\times E_{8}$ heterotic string compactified on a Calabi-Yau threefold with a fibration by special Lagrangian 3-tori, known as an SYZ fibration \cite{Strominger:1996it}. The conditions of $N=1$ supersymmetry additionally require that the heterotic background gauge field must satisfy the Hermitian Yang-Mills equations. The equations we study in this paper may be understood as an approach to this duality from the heterotic side, where they give the lowest order $\a$-corrections to the heterotic large volume (i.e. M-theory half-K3) limit. This is a step towards the $\a$-corrected heterotic dual of Donaldson's local adiabatic limit of co-associative fibered $G_2$ manifolds \cite{donaldson2017adiabatic}, applicable in the M-theory setting. 

Compact spaces of the required type for physically realistic effective theories are not yet available: on the M-theory side, no compact $G_{2}$ holonomy spaces with codimension-7 singularities sitting inside orbifold loci have been constructed, and on the heterotic side, Hermitian Yang-Mills connections over compact SYZ fibrations are not well-understood. Thus we are currently limited to working with local models of such geometries. The Hull-Strominger system, when reduced on the fibers of a local model $T^{3}\times\mathbb{R}^{3}$ of the SYZ geometry, gives an $\a$-corrected version of the equations for a stable complex flat connection in 3D, which were introduced in \cite{Acharya:2002kv}. These equations give a fruitful playground for understanding the matter spectrum of $G_2$ (and heterotic) compactifications, and have been studied in various ways in the literature. A method for computing spectra of solutions via Morse cohomology was given in \cite{Pantev:2009de}, and the method was extended and applied in \cite{Braun:2018vhk} to reduced models of twisted connected sum $G_2$ holonomy spaces \cite{kovalev,Corti:2012kd,Corti2013}. The cohomology method was further extended to local $G_2$ holonomy spaces in \cite{Hubner:2020yde}. The first non-abelian solution was given in \cite{Barbosa:2019bgh}, where an $SU(3)$ solution was constructed from the M-theory perspective via T-branes \cite{Cecotti:2010bp}. The authors of \cite{Barbosa:2019bgh} point out that such non-abelian solutions can allow for chiral zero-modes, even if the upstairs $G_2$ geometry locally has only a codimension six singularity, and they present a local example with an explicit construction of such a chiral mode. 

In this paper, we study non-abelian stable complex flat connections from a heterotic perspective. We present a Morse-Witten type cohomology that can be used to compute the index of such solutions. If we make some additional assumptions for the types of non-abelian solutions considered, we can also use this cohomology to compute the spectrum. We also investigate how $\a$-corrections coming from the local reduced Hull-Strominger system correct and change the tree-level solutions. We consider how $\a$-corrections may modify the generic behaviour of solutions near sources, and how the metric and D-term stability condition receive non-trivial corrections due to the heterotic Bianchi-Identity close to the sources. In particular, for generic non-abelian solutions, we are forced to introduce a two-form field which alters the equations in an interesting way.

The paper is organised as follows. In section 2, we introduce the Hull-Strominger system, reduce it on a local model for an SYZ-fibered Calabi-Yau threefold, and compute the $\a$-corrections to the resulting 3D equations for a stable complex flat connection. In section 3, we explore solutions to this system of equations, beginning with abelian solutions and then examining a non-abelian monopole-type solution. In section 4, we introduce a method for computing the chiral spectrum or chiral index of non-abelian solutions that become asymptotically abelian near sources of the Higgs field, or at least remain well-behaved. In section 5, we present examples of spectrum computations. In section 6, we give our conclusions and further directions. 

\section{The Hull-Strominger System}
We start by first recalling the $\a$-corrected system of equations that must be satisfied by the geometry, gauge bundle, dilaton, and B-field of an $N=1$ heterotic background. The manifolds of interest admit an $SU(3)$ structure $(X,\Omega,\omega)$, where $\Omega$ is a complex nowhere vanishing three-form and $\omega$ is a real two-form of maximal rank. The form $\Omega$ is also required to be locally decomposable, which implies that it endows $X$ with an almost complex structure $J$ \cite{Hitchin2001}. Moreover, $\Omega$ is a $(3,0)$ form with respect to $J$. 

The forms $\Omega$ and $\omega$ now satisfy the usual $SU(3)$ structure relations
\begin{equation}
\label{eq:su3}
\tfrac{i}{\vert\vert\Omega\vert\vert^2}\Omega\wedge\bar\Omega=\tfrac16\omega\wedge\omega\wedge\omega\:,\;\;\;\omega\wedge\Omega=0\:.
\end{equation}
The first equation identifies the volume forms defined by the two structure forms, while the second relation says that $\omega$ is of type $(1,1)$ with respect to $J$. For supersymmetric heterotic compactifications to four dimensional Minkowski space,
\begin{equation}
{\cal M}_{9,1}={\cal M}_{3,1}\times X\:,
\end{equation}
the internal geometry $X$ is required to satisfy the relations \cite{Strominger1986, Hull1986357, Gauntlett:2003cy}
\begin{align}
\label{eq:SUSY1}
\d(e^{-2\phi}\Omega)&=0\\
\label{eq:SUSY2}
\d(e^{-2\phi}\omega\wedge\omega)&=0\\
\label{eq:SUSY3}
i(\bp-\p)\omega&=H\:,
\end{align}
where $\phi$ denotes the heterotic dilaton and $H$ is given by 
\begin{equation}
\label{eq:Hflux}
H=\d B + \tfrac{\a}{4}\left(\omega_{CS}(A) - \omega_{CS}(\nabla)\right)\:.
\end{equation}
Here $A$ is the gauge connection of a vector bundle with structure group contained in either $E_8\times E_8$ or $SO(32)$, while $\nabla$ is an $\End(TX)$ valued connection which will play less of a role for us\footnote{The freedom to choose $\nabla$ has been discussed extensively in the literature, see e.g. \cite{Hull198651, Hull1986187, Sen1986289, Ivanov:2009rh, Martelli:2010jx, delaOssa:2014msa}.}. The two-form $B$ is the heterotic Kalb-Ramond field, which has to transform under gauge transformations in order that the flux $H$ remains gauge invariant \cite{Green1984117}. The three form then satisfies the heterotic Bianchi identity
\begin{equation}
\label{eq:BI}
\d H=2i\p\bp\omega=\tfrac{\a}{4}\left(\tr\:F\wedge F - \tr\: R\wedge R\right)\:,
\end{equation}
where $F$ and $R$ are the curvatures of $A$ and $\nabla$ respectively. For our local explicit solutions, the last term on the right hand side will be of cubic order in $\a$ and will hence be dropped from now on. However, this term will be important when one considers compact global issues, and we will investigate this further in future publications.

The first condition \eqref{eq:SUSY1} implies that the complex structure defined by $\Omega$ is {\it integrable}, so $(X,J)$ is a complex manifold. The second condition \eqref{eq:SUSY2} is known as the conformally balanced condition, while the third condition \eqref{eq:SUSY3} identifies the heterotic NS three-form flux in terms of the internal geometry. In addition, the gauge bundle is required to satisfy the supersymmetry conditions
\begin{equation}
\label{eq:SUSY4}
F\wedge\Omega=0\:,\;\;\;\omega\wedge\omega\wedge F=0\:,
\end{equation}
often referred to as the Hermitian Yang-Mills conditions. Indeed the first condition implies that $F$ is of type $(1,1)$, which means that the bundle is {\it holomorphic}. The second condition is the Yang-Mills constraint, which implies that the bundle is poly-stable by the Donaldson-Uhlenbeck-Yau theorem \cite{donaldson1985anti, uhlenbeck1986existence}.

The system of equations \eqref{eq:SUSY1}-\eqref{eq:SUSY3} together with the heterotic Bianchi identity \eqref{eq:BI} and the Hermitian Yang-Mills conditions \eqref{eq:SUSY4} are often referred to as the Hull-Strominger system. It is (perturbatively) accurate modulo cubic corrections in $\a$.

\subsection{Reducing the Hull-Strominger system on $T^3\times M_3$}
We are interested in solutions of the Hull-Strominger system on local models for Calabi-Yau manifolds, or more generally torsional models solving the Hull-Strominger system. In particular, we assume that the internal geometry has a special Lagrangian $T^3$ fibration \cite{Strominger:1996it}. Locally we can model such a geometry as
\begin{equation}
\label{eq:localMod}
X=T^3\times M_3\:.
\end{equation}
We will also assume the fibers $T^3$ to be sufficiently small, so that we can have our solution depend nontrivially only on the $M_3$ coordinates. In this paper we will take $M_3$ to be either $S^3$ or $\mathbb{R}^3$ for local models of $M_3$. Let us proceed to consider such local models where $M_3=\mathbb{R}^3$.

The tree-level ($\alpha'=0$) geometry then consists of a Calabi-Yau metric on this space, with a local complex structure given by
\begin{equation}
\label{eq:CplxStr}
\d z^i=\d x^i+i\d\phi^i\:,
\end{equation}
where $\{x^1,x^2,x^3\}$ are the coordinates on $M_3$, while $\{\phi^1,\phi^2,\phi^3\}$ are the coordinates of the three-torus. Other ansatze for the geometry are of course possible. In particular, we discuss an ansatz in appendix \ref{app:T'Hooft-Polyakov geometry} with relevance for local heterotic/F-theory duality, which also turns out to result in a $\a$-corrected version of the t'Hooft-Polyakov monopole \cite{t1974magnetic, polyakov1996particle}.

Locally on $T^3\times\mathbb{R}^3$ we have the corresponding complex top-form and Hermitian form given by
\begin{equation}
\label{eq:ZeroG}
\Omega_0=\d z^1\wedge \d z^2\wedge \d z^3\:,\;\;\;\;\omega_0=\lambda^2\sum_{i=1}^3\d x^i\wedge\d\phi^i={g_0}_{ij}\d x^i\wedge\d\phi^j\:,
\end{equation}
where $\lambda$ denotes a constant size parameter. This model $SU(3)$ structure corresponds to a flat tree-level metric. We will use this as our model for explicit $\mathbb{R}^3$ computations throughout this paper, leaving reductions to more generic curved backgrounds for future work.

Let us then reduce the Hermitian Yang-Mills equations on this system. We first expand
\begin{equation}
A=A^x_i\d x^i+A^\phi_i\d\phi^i\:,
\end{equation}
where $A^x$ and $A^\phi$ are assumed to depend only on the non-compact coordinates. We have
\begin{equation}
\d_A=\d+\,A\:,\;\;\;F=\d A+\,A\wedge A\:.
\end{equation}
Plugging this into the holomorphic constraint gives
\begin{align}
\label{eq:Flat1}
F^x&=\,A^\phi\wedge A^\phi\\
\label{eq:Flat2}
\d_{A^x}A^\phi&=0\:,
\end{align}
where $d_{A^x}$ is the exterior covariant derivative with respect to the connection $A^x$, and where $A^\phi = A^\phi_i\,\d x^i$ transforms as a one-form on $\mathbb{R}^3$ due to a topological twist. This twisting occurs because the $T^3$ fiber is a special Lagrangian, so that its normal bundle is isomorphic to its tangent bundle \cite{Acharya:1998pm}. The F-term equations \eqref{eq:Flat1} and \eqref{eq:Flat2} can be recast as the equation for a complex flat connection
\begin{equation}
\label{eq:CpxFlat}
{\cal A}=A^x+i\,A^\phi\:,\;\;\;F_{\cal A}=\d{\cal A}+\A\wedge\A=0\:.
\end{equation}
Reducing the Yang-Mills conditions gives a co-closure constraint on $A^\phi$. At zeroth order in $\a$, the equation reads
\begin{equation}
\label{eq:CoclosureZero}
d^\dagger_{A^x} A^\phi=0\:.
\end{equation}
where the dagger denotes an adjoint taken with respect to the tree-level metric from \eqref{eq:ZeroG}. This equation can be viewed as a stability condition on the flat connection \cite{corlette1988flat}. Together, the F-term equation \eqref{eq:CpxFlat} and D-term equation \eqref{eq:CoclosureZero} comprise the equations for a stable complex flat connection in 3D.

Including $\a$-corrections modifies the D-term co-closure equation while preserving the F-term flatness condition, as we will see in the next subsection. 

\subsection{The Back-Reacted Geometry}
Now we will consider $\a$-corrections to our equations. Instanton configurations of the type \eqref{eq:Flat1}-\eqref{eq:Flat3} will back-react on the geometry through the heterotic Bianchi identity \eqref{eq:BI} and supersymmetry conditions \eqref{eq:SUSY1}-\eqref{eq:SUSY3}. Such back-reactions will only become relevant close to sources, and so we restrict ourselves to a local patch containing the source. 

A short computation using \eqref{eq:Flat1}-\eqref{eq:Flat2} reveals that for the reduced geometry, the Pontryagin class may be written as
\begin{equation}
\tr\,F\wedge F=-\d_x\tilde\d_x\tr\left (A^\phi\wedge\tilde A^\phi\right)+\tfrac{2}{3}\,\d_x\tr\left(\tilde A^\phi\wedge \tilde A^\phi\wedge \tilde A^\phi\right)-\tfrac{2}{3}\,\tilde\d_x\tr\left(A^\phi\wedge A^\phi\wedge A^\phi \right)\:,
\end{equation}
where $\d_x=\d x^i\partial_i$, $\tilde\d_x=\d \phi^i\partial_i$ and $\tilde A^\phi=\d\phi^i A^\phi_i$. Using the Hodge decomposition of $M_3$, it follows that
\begin{equation}
\tfrac{2}{3} \,\tilde\d_x\tr\left(A^\phi\wedge A^\phi\wedge A^\phi \right)=\tilde\d_x\d_x B
\end{equation}
for some real two-form $B\in\Omega^2(M_3)$. We then find
\begin{equation}
\tr\,F\wedge F=\d_x\tilde\d_x\Big(B+\tilde B-\tr\left (A^\phi\wedge\tilde A^\phi\right)\Big)\:,
\end{equation}
where $\tilde B=\tfrac12B_{ij}\d \phi^{ij}$. From the Bianchi identity \eqref{eq:BI}, it follows that $\omega$ is corrected to 
\begin{equation}
\label{eq:omega1}
\omega=\omega_0 - \tfrac{\a}{16}\Big(\tr\left(A^\phi\wedge\tilde A^\phi\right)-B-\tilde B\Big)\:,
\end{equation}
where $\omega_0$ is the tree-level solution, which satisfies
\begin{equation}
\d_x\tilde\d_x\omega_0=0\:.
\end{equation}
Locally on $\mathbb{R}^3$ we will take $\omega_0$ to be given by \eqref{eq:ZeroG} above, i.e. corresponding to the flat metric. We will consider situations where the zeroth order geometry is also curved in a later publication. 
From $\omega$ we get a corrected metric
\begin{equation}
\label{eq:RedHSMetric}
g_{mn}={g_0}_{mn} - \tfrac{\a}{16}\tr\left(A^\phi_m A^\phi_n\right)\:,
\end{equation}
on $M_3$, where $g_0$ is the metric derived form $\omega_0$.

Before we consider the $\a$-correction to \eqref{eq:CoclosureZero}, let us first make an observation concerning the two-form $B$ we introduced above. It is tempting to speculate that doing an appropriate $\a$-correction to the complex structure ansatz \eqref{eq:CplxStr}, we can absorb $B$, leaving a Hermitian form as in \eqref{eq:ZeroG} but with ${g_0}_{ij}=\lambda^2\delta_{ij}$ replaced by $g_{ij}$. However, it turns out that if the deformed complex structure is to remain integrable, then $B$ must be a closed two-form if it is to be cancelled by such a deformation. As only the exterior derivative of $B$ appears in the Bianchi identity, we see that such a deformation cannot absorb $B$. 

The $\a$-corrections to the D-term supersymmetry equation, which is the reduction of the Yang-Mills condition for the gauge sector, reads
\begin{equation}
\label{eq:Flat3}
g^{mn}\nabla^x_mA_n^\phi=\tfrac{\a}{16}B^{nq}A_n^\phi A_q^\phi\:,
\end{equation}
where indices are raised and lowered using $g_{mn}$, and $\nabla^x_m$ acts as the Levi-Civita connection on space-time indices. Note at zeroth order in $\a$ this is just the co-closure condition, but for general non-abelian connections where $B\neq0$ the equation gets corrected even at first order in $\a$. The appearance of the Levi-Civita connection for the $\a$-corrected metric does not come directly from reduction of the Hermitian Yang-Mills equations, but is implied by the topological twist discussed in the previous subsection. Interestingly, the additional terms in the covariant derivative required by the Levi-Civita connection vanish to first order in $\a$ when contracted with the inverse metric, so the different choices of connection are equivalent to the order to which the Hull-Strominger system is corrected. 

We also need to check the conformally balanced condition \eqref{eq:SUSY2}. A somewhat lengthy computation gives the dilaton factor as
\begin{equation}
\d_x\phi=\tfrac{\a}{64}\d_x\left(\tr( A_m^\phi{A^\phi}^m)+\tfrac{\a}{32}\tr(A_m^\phi{A_n^\phi})\tr({A^\phi}^m{A^\phi}^n)+\tfrac{\a}{32}\kappa\,f-\tfrac{\a}{32}B_{mn}B^{mn}\right)\:,
\end{equation}
where we have used \eqref{eq:Flat3}, and where one must also choose $B$ so that 
\begin{equation}
\nabla^mB_{mn}=0\:.
\end{equation}
Here indices are again raised and the adjoint is taken with respect to the metric $g$. We are free to assume $B$ is co-exact, given in terms of a function $f$ as
\begin{equation}
B_{np}=\nabla^m(f\,\epsilon_{mnp})\:.
\end{equation}
The constant $\kappa$ is the part of $\tr({A^\phi}\wedge{A^\phi}\wedge{A^\phi})$ proportional to the volume form
\begin{equation}
\tfrac{2}{3}\,\tr({A^\phi}\wedge{A^\phi}\wedge{A^\phi})=\d_xB+\kappa*1\:.
\end{equation}
On $\mathbb{R}^3$ this can be absorbed in $\d_xB$, as any three-form on $\mathbb{R}^3$ is exact by the Poincare lemma. 

Finally, we can also check that \eqref{eq:SUSY1} is satisfied, provided that we take
\begin{equation}
\Omega=e^{2\phi}\Omega_0\:.
\end{equation}
Additionally, it can be verified that the $SU(3)$ conditions \eqref{eq:su3} are also satisfied.

\section{Solutions to the Reduced System}

The natural place to begin a search for stable complex flat connections is smooth solutions of finite energy. However, in analogy to the 4D Hitchin system on $\mathbb{R}^{4}$, it seems likely that any smooth, finite energy stable complex flat connection is gauge equivalent to the trivial solution. We will show below that this is true if one assumes strong enough falloff conditions on the Higgs field. The implications are that finite energy solutions must not be smooth, which we take to mean that the Higgs field has singularities along a configuration of sources. 

To see that the claim is true for strong enough falloff conditions,
we look at a corollary of the Bochner-Weitzenboch identity that holds for solutions
of the complex Yang-Mills equations (and thus for solutions of \eqref{eq:CpxFlat}-\eqref{eq:CoclosureZero}):
\[
\frac{1}{2}\Delta\left|A^{\phi}\right|^{2}-\left|\nabla_{A^{x}}A^{\phi}\right|^{2}-\left| A^{\phi} \wedge A^{\phi}\right|^{2}=0
\]
where $A^{x}+iA^{\phi}$ satisfies the complex Yang-Mills equations
and we have specialized to $\mathbb{R}^{3}$ to drop a term involving
the Ricci curvature \cite{gagliardo2012geometric}. We then integrate this equation over $\mathbb{R}^{3}$.
The Laplacian term may be integrated via Stoke's theorem, and with
strong enough falloff conditions on $A^{\phi}$, this tells us that
the integral of this term vanishes. In that case, we have that the
sum of the other two terms vanishes. Because these terms are negative
semidefinite, we conclude that 
\[
\nabla_{A^{x}}A^{\phi}=0,\quad A^{\phi} \wedge A^{\phi}=0
\]
The second equation tells us that $F^x$ is zero, via the F-term equation, and because we are on $\mathbb{R}^{3}$, we may then choose a gauge in which $\nabla_{A^{x}}=\d$, and then the above equations imply that $A^{\phi}$ is trivial. 

It seems likely that the 6D finite energy condition, when reduced to
3D, will provide strong enough falloff conditions for the vanishing
of the integrated Laplacian above. To prove this hypothesis would
require additional analysis of 3D gauge theory that we leave for future
work. 

The above argument applies only to the tree level $\a=0$ equations. For the $\a$-corrected system, we do not have access to a Bochner-Weitzenboch identity, so cannot rule out smooth solutions in the same way. However, in the cases examined below, including $\a$-corrections to a solution that is singular at tree level does not smooth out the singularity, and instead worsens the singularities (at least at $O(a')$, where the solution is reliable). 

\subsection{Abelian Solutions}
The simplest solutions to the reduced Hull-Strominger system are found by taking the gauge group to be abelian. It turns out that, as we will see below, these are also relevant to certain non-abelian solutions with sources. Indeed, we will argue that for a certain type of non-abelian solution, the part of the connection which sees the source effectively becomes abelian. It thus makes sense to investigate closer the local geometry of such solutions near sources. These solutions were also extensively studied as local models for M-theory compactifications on $G_2$ manifolds in \cite{Acharya:2001gy, Pantev:2009de, Braun:2018vhk, Barbosa:2019bgh, Barbosa:2019hts}.

The Higgs field of an abelian solution satisfies $A^\phi \wedge A^\phi = 0$, which by \eqref{eq:Flat1} implies $F^x = 0$. Assuming we are on a simply connected space $M_3$, we may then choose $A^x=0$.\footnote{Our space will either be $S^3$ or the local model $\mathbb{R}^3$.} Thus we can decouple the gauge and tangent bundle factors of $A^\phi$ and search for $A^\phi\in\Omega^1(M_3)$ satisfying the F-term and D-term equations
\begin{equation}
\d A^\phi=\d^{\dagger} A^\phi=0\:,
\end{equation}
where the adjoint is taken with respect to the $\a$-corrected metric. Thus we are looking for harmonic 1-forms on $M_3$. We will work in a patch, so that the closure condition on $A^\phi$ implies by the Poincare lemma that we can set
\begin{equation}
A^\phi=\d\psi\:,
\end{equation}
for some real function $\psi$. The co-closure condition then becomes
\begin{equation}
\label{eq:coclose}
\d^{\dagger}\d\psi=0\:,
\end{equation}
and so we are looking for harmonic functions on $\mathbb{R}^3$, which may blow up at sources. Note that the metric is given as
\begin{equation}
g_{mn}={g_0}_{mn}-\tfrac{\a}{8}\,\partial_m\psi \partial_n\psi\:,
\end{equation}
where the factor of $2$ relative to \eqref{eq:RedHSMetric} comes from the trace, and the form of $g_0$ is given by \eqref{eq:ZeroG}. The dilaton is
\begin{equation}
\phi=\tfrac{\a}{64}\,g^{mn}\partial_m\psi \partial_n\psi +\tfrac{\a^2}{32\times 64}\left(g^{mn}\partial_m\psi \partial_n\psi\right)^2\:,
\end{equation}
where we have set the constant part of the dilaton to zero. Let's go on to consider some common local solutions on $\mathbb{R}^3$ and their $\a$ corrections.

At zeroth order, the radially symmetric solution is the standard monopole harmonic function
\begin{equation}
\label{eq:monopole}
\psi(r)=\tfrac{C}{r}+{\cal O}(\a)\:,
\end{equation}
for some constant $C$ corresponding to the charge of the monopole. Given the monopole solution \eqref{eq:monopole} for $\psi$, we find that the dilaton is
\begin{equation}
\phi=\frac{\a\,C^2}{64\lambda^2\,r^4}+{\cal O}(\a^2)\:.
\end{equation}
The only component of the metric that is corrected is $g_{rr}$:
\begin{equation}
g_{rr}=\lambda^2+\tfrac{\a C^2}{8 r^4}+{\cal O}(\a^2)\:.
\end{equation}
Note that both the dilaton and metric blow up as $r\rightarrow 0$. 

\subsubsection*{Exact Solution and large charge limit}
Local solutions to torsional heterotic compactifications and the Hull-Strominger system with abelian bundles have been studied from differrent perspectives before \cite{Carlevaro:2008qf, Carlevaro:2009jx, Halmagyi:2016pqu, Halmagyi:2017lqm}. The benefits of studying abelian bundles is that a particular double scaling limit can be employed, where the charge of the gauge field is sent to infinity, while $\a$ is sent to zero in a controlled manner. This results in a finite correction to the geometry at first order in $\a$, while higher corrections vanish. The solution is one-loop exact.

We consider the exact solution to the reduced Hull-Strominger system with a radially-symmetric potential field $\psi(r)$. If we solve equation \eqref{eq:Flat3} using the $\a$-corrected metric, we find
\begin{equation}
\psi'(r)=-\tfrac{C}{\sqrt{r^4-\a\tfrac{C^2}{8 \lambda^2}}} 
\end{equation}
This Higgs field no longer blows up at the origin, but at a non-zero radius. From this Higgs field, we can calculate the metric, which again is corrected only in its $g_{rr}$ component:
\begin{equation}
\label{eq:ExactAbMetric}
g_{rr}=\tfrac{\lambda^2}{1-\a \tfrac{C^2}{8 \lambda^2 r^4}}\:.
\end{equation}
This Higgs field, metric, and Riemann curvature blow up at the finite radius $r_0^4 = \tfrac{\a C^2}{8 \lambda^2}$, indicating that there is a spherical source at this radius. The Higgs field becomes imaginary inside $r=r_0$, so our solution is unphysical inside this radius and cannot tell us about the interior. The radius $r_0$ depends on the size parameter $\lambda$ such that in the large volume limit $\lambda \to \infty$, the singularity becomes concentrated near the origin. We also see from \eqref{eq:ExactAbMetric} that the appropriate double scaling limit to consider when sending $\a$ to zero is to rescale the charge so that $\a C^2$ remains finite. 

There are no further corrections to the Higgs field or metric from the Hull-Strominger system. However, outside of the large charge limit the D-term equation itself is expected to receive further corrections at $O(\a ^2)$, so the exact solution to the $O(\a)$ equation may not be physically reliable at higher orders in the SUGRA expansion. One may also extend this analysis to the case of multiple monopole sources in a straightforward way. 

We can also consider the solution close to a one-dimensional source. This is effectively a two-dimensional problem with cylindrical symmetry. The zeroth order solution now reads
\begin{equation}
\psi(\rho)=C\log(\rho)+{\cal O}(\a)\:,
\end{equation}
where now $C$ denotes the charge density along the source. The results for this case are very similar to the monopole case, but with the substitution $r^2 \rightarrow \rho$. In particular, the fully corrected Higgs field and metric are now
\begin{equation}
\psi'(r)=-\tfrac{C}{\sqrt{\rho^2 -\a \tfrac{C^2}{8 \lambda^2}}}
\end{equation}
and
\begin{equation}
g_{\rho \rho}= \tfrac{\lambda^2}{1-\a \tfrac{C^2}{8\lambda^2 \rho^2}} \:
\end{equation}
This solution blows up at a nonzero radius in the plane transverse to $\rho = 0$, indicating a cylindrical source surrounding this line. 

One may also examine solutions of the reduced Hull-Strominger system with negative $\a$, which have interesting behaviors, though their physical relevance is less clear. In this case, for the radially-symmetric solution, we find that the Higgs field is everywhere smooth, while the metric has a curvature singularity at the origin, but is smooth at $r=r_0$. We may imagine smoothly adjusting $\a$ from a positive to a negative value and tracking the behavior of the Higgs field singularity along the way: for $\a>0$, there is a singular horizon at $r=r_0(\a)$, which contracts as we decrease $\a$. When $\a=0$, the singularity sits at the origin in the Higgs field only, and when we continue to $\a<0$, the singularity moves instead to the metric only, where it becomes a curvature singularity at the origin. 

\subsection{$SU(N)$ Solutions}

Let us now look for solutions to the reduced Hull-Strominger system for non-abelian gauge fields. In particular, we will restrict ourselves to gauge group $SU(N)$, and our main example will have an $SU(2)$ gauge group. We will first consider the configurations at tree-level in $\a$, and then discuss $\a$-corrections at the end of the section. 

Our complex connection on $\mathbb{R}^3$ is given by
\begin{equation}
\d_{A^x}=\d+ \,A^x\:,
\end{equation}
where $A^x$ is a one-form valued in the Lie-algebra of $SU(2)$, i.e. the span of the anti-Hermitian Pauli matrices (in math conventions). A complex connection $\A$ can now be constructed as
\begin{equation}
\A=A^x+i\,A^\phi\:,
\end{equation}
where we require both $A^x$ and $A^\phi$ to be anti-Hermitian, valued in ${\mathfrak{su}}(2)$ with legs now on the three-dimensional base. This implies that $\A$ takes values in the Lie algebra of $SL(2,\mathbb{C})$. The flatness condition on $\cal A$ then implies that
\begin{equation}
\label{eq:FlatA}
\A= \,G^{-1}\d G\:,
\end{equation}
where $G\in SL(N,\mathbb{C})$, at least locally. Because we are working on a local model, we may take \eqref{eq:FlatA} as our ansatz. We thus have the real and imaginary parts of $\cal A$ given by
\begin{align}
A^x&=\tfrac{1}{2}\left(G^{-1}\d G-\d G^\dagger(G^\dagger)^{-1}\right)\\
A^\phi&=-\tfrac{i}{2}\left(G^{-1}\d G+\d G^\dagger(G^\dagger)^{-1}\right)\:.
\end{align}

We now use the polar decomposition which states that any invertible matrix can be represented as
\begin{equation}
G=HU\:,
\end{equation}
where $U$ is unitary and $H$ is Hermitian matrix of positive eigenvalues. When $G$ has unit determinant, which we will assume, both $H$ and $U$ may be chosen to have unit determinant as well, so that $U\in SU(N)$. A transformation
\begin{equation}
G\rightarrow G\tilde U\:,
\end{equation}
where $\tilde U\in SU(N)$ transforms the gauge field and Higgs field as
\begin{align}
A^x&\rightarrow \tilde U^\dagger A^x\tilde U + \tilde U^\dagger\d\tilde U\\
A^\phi&\rightarrow \tilde U^\dagger A^\phi\tilde U\:.
\end{align}
This then corresponds to usual gauge transformations. We can use this to make $G$ Hermitian, since there is always a gauge where
\begin{equation}
G=H\:.
\end{equation}
In this gauge, the gauge field and Higgs field read
\begin{align}
A^x&=\tfrac{1}{2}[H^{-1},\d H]\\
A^\phi&=-\tfrac{i}{2}\{H^{-1},\d H\}\:,
\end{align}
where curly brackets denote the anti-commutator. An anti-Hermitian matrix of unit determinant has $N^2-1$ degrees of freedom (the dimension of $SU(N))$. The co-closure condition
\begin{equation}
d_{A^x}^\dagger A^\phi=0
\end{equation}
then gives a non-linear second order differential equation to be solved for $H$.

\subsection*{$SU(2)$ Solutions}

Now let us specialize to $N=2$. We may further parameterise $H$ as
\begin{equation}
H=H_0+c\,I_2\:,
\end{equation}
where $H_0$ is traceless, $c\in\mathbb{R}$, and $I_2$ is the rank-2 identity matrix. We expand $H_0$ in Pauli matrices as 
\begin{equation}
H_0=\sum_ia_i\sigma_i\:,
\end{equation}
The condition that $H$ has unit determinant is then
\begin{equation}
\label{eq:DetH}
c^2-\sum_ia_ia_i=1\:.
\end{equation}
We are hence left with an overall number of four parameters $\{c,a_1,a_2,a_3\}$ describing the complex flat connection, subject to the constraint \eqref{eq:DetH}. 

In terms of $H$, the one forms $A^x$ and $A^\phi$ now read
\begin{align}
\label{eq:AxH}
A^x&=-\tfrac{1}{2}[H_0,\d H_0]\\
\label{eq:ApH}
A^\phi&=i\left(H_0\d c-c\,\d H_0 \right)\:,
\end{align}
where we have used the relation \eqref{eq:DetH} which implies that
\begin{equation}
\d(H_0^2)=\d(c^2)I_2\:.
\end{equation}
We now come to the tree-level stability equation
\begin{equation}
\label{eq:StabH}
\p_iA^\phi_i+[A^x_i,A^\phi_i]=0\:.
\end{equation}
Plugging in \eqref{eq:AxH} and \eqref{eq:ApH} into this equation, we find the following nonlinear differential equation
\begin{equation}
-c^2\Delta a_m+(\Delta c)a_m+2a^2\p_ic\p_ia_m-\p_ic\p_i(a^2)a_m-c\p_i(a^2)\p_ia_m+2c(\p_ia_j)(\p_ia_j)a_m=0\:.
\end{equation}
where $a^2=a_ja_j$ and $\Delta=\p_i\p_i$. Multiplying by $c$ and using \eqref{eq:DetH}, this can be simplified a bit to
\begin{equation}
\label{eq:NonlinODE}
-c^2\Delta a_m+(c\Delta c)a_m-\tfrac12\p_i(a^2)\p_i(a^2)a_m-\p_i(a^2)\p_ia_m+2c^2(\p_ia_j)(\p_ia_j)a_m=0\:.
\end{equation}
Recall that the function $c$ is determined by the $a_i$'s through \eqref{eq:DetH}. 

This is a rather complicated nonlinear differential equation, and it is not practical to find a general solution. One might try to simplify matters by, for example, assuming that the field $c$ can be taken to be constant. This leads to the simpler equation
\begin{equation}
\Delta a_m-2(\p_ia_j)(\p_ia_j)a_m=0\:,
\end{equation}
where by \eqref{eq:DetH} we have used that $a^2$ will also be constant in this case. Contracting this equation by $a_m$, and using \eqref{eq:DetH} agaain, we find
\begin{equation}
(\p_ia_j)(\p_ia_j)(2a^2+1)=0\:,
\end{equation}
which can only be satisfied if $\p_ia=0$. We conclude that in order to have nontrivial solutions to \eqref{eq:NonlinODE}, we need $\p_ic\neq 0$.

\subsection{Monopole-Type Solution}
\label{sec:ExamplesFlat}
We would like to find a Hermitian matrix that satisfies the constraint \eqref{eq:DetH} and solves equation \eqref{eq:StabH}. As we have seen, the general equation is quite complicated, but the hope is that we can find a clever parameterisation of $H$ which solves the system. To get a foothold, let's consider again \eqref{eq:DetH}. A convenient parameterisation of $c$ and $a^2=a_ia_i$ then reads
\begin{align}
c&=\cosh(u(x^i))\\
a^2&=\sinh(u(x^i))\:,
\end{align}
for some function $u(x^i)$. We will assume that $u(x^i)$ is radially symmetric, so that it depends only on the radius $r$. We then write
\begin{equation}
H_0=\sinh(u(r))\tilde a_i\sigma_i\:,
\end{equation}
where we have introduced the normalized $\tilde a_i$ so that $\tilde a_i\tilde a_i=1$. We want to allow the $\tilde a_i$ to depend on coordinates other than $r$, since otherwise the field $A^\phi$ will square to zero and the curvature is flat everywhere. We may try a simple ansatz inspired by the t'Hooft-Polyakov monopole \cite{t1974magnetic, polyakov1996particle}:
\begin{equation}
\tilde a_i=\hat x^i=\tfrac{x_i}{r}\:.
\end{equation}
In this case, the eigenvalues of $H$ are given by $e^{\pm u}$.

Plugging this ansatz into the equations and solving the system using Mathematica, we find that the reduced Strominger system is indeed solved provided the function $u$ satisfies the equation
\begin{equation}
\label{eq:RadExLapl}
2r^2\,\Delta u(r) = \sinh(4u(r)) \:.
\end{equation}
where $\Delta$ is the Laplacian. Equation \eqref{eq:RadExLapl} becomes even simpler if we view it in terms of the inverse variable
\begin{equation}
t=\tfrac{1}{r}\:.
\end{equation}
We get that
\begin{equation}
\label{eq:RadialEq2}
2t^2\,u''(t) = \sinh(4u(t)) \:,
\end{equation}
Note that this equation implies that the second derivative of $u(t)$ always takes the sign of $u(t)$. In particular, a solution that tends to zero at large $r$ must necessarily blow up at some small $r$. To see this, assume that $u(t)\rightarrow0$ as $t\rightarrow0$. Then the equation to first order in $t$ is
\begin{equation}
4u(t) - 2t^2\,u''(t)=0\:,
\end{equation}
with solution
\begin{equation}
\label{eq:linearApprox}
u(t) = C_1\,t^2  + \tfrac{C_2}{t}\:.
\end{equation}
To avoid a singularity at $t=0$, we must set $C_2=0$. The remaining solution will continue to grow for larger $t$, i.e. as $r$ tends to zero.

Equation \eqref{eq:linearApprox} is a good approximation for the solution in a region of $t$ when $u(t)$ is small, but the non-linear effects in \eqref{eq:RadialEq2} from the hyperbolic sine will sooner or later come into play. Numerical results suggest that these non-linear effects force the solution to blow up at finite $t$. Thus any solution that tends to zero at $t=0$ will at best be defined on an interval of the form $(0,t_1)$, while a solution that tends to zero at $t=\infty$ is at best defined on an interval $(t_1,\infty)$, where the solution blows up at $t_1>0$. Switching back to the $r$ coordinate, It is also interesting to note that for solutions that tend to zero at the origin, we have $u$ growing linearly in $r$ away from zero. Hence, the eigenvalues of $H$ are not smooth at the origin for such solutions, although it can be checked that the complex flat connection is nonetheless well-defined and smooth. 

We have sources where $u(r)\rightarrow\pm\infty$, meaning that the sources are spherical, as in the earlier $\a$-corrected abelian example. We will consider later what happens to the solution as we approach such sources. 

\subsubsection*{$\a$-Corrections}
Now we will add the $\a$-corrections to the monopole-type solution. To do so, we keep the same ansatz for the complex flat connection, but we modify our D-term equation to \eqref{eq:Flat3}. Again, this matrix equation reduces to an ODE for the function $u(r)$: 
\begin{equation}
\label{eq:monoDterm}
\Delta_{\tilde{g}} u(r) = \frac{512r^2 \sinh (4u(r)) + 8\a \sinh^2(2u(r))\bigr( 4u(r)+\sinh(4u(r)) \bigr)}{\bigr( 32r^2 + \a \sinh^2(2u(r)) \bigr)^2} \ ,
\end{equation}
where $\Delta_{\tilde{g}}$ is the Laplacian with respect to the $\a$-corrected metric defined in \eqref{eq:RedHSMetric}. 

Unlike the abelian case it is difficult to define a large-charge limit. The $\a$-corrections are generically not one-loop exact, and we find qualitatively different results when solving the equations to $\cal{O}(\a)$ or exactly in $\a$. This is as expected, because the Hull-Strominger system includes only the first order correction in $\a$, and higher order effects are expected to enter from supergravity and gauge theory sectors beginning at ${\cal O}(\a^2)$.

Numerical solutions of the equation truncated to first order in $\a$ suggest that the singularity behavior for solutions to the one-loop equation are similar to that of the uncorrected equation, with generic solutions existing on an interval $(r_1,r_2)$. Meanwhile, numerical solutions to the exact D-term equation have no singularities in $u(r)$ or its first derivative for finite $r$, although $u(r)$ blows up linearly as $r\to\infty$. The solution approximates the zeroth order solution and $\a$-corrected solution well in the interval $(r_1,r_2)$. Thus, when compared to the $\a=0$ solution, the exact solution to the $\a$-corrected equation seems to smear the non-abelian sources such that the finite-$r$ singularities disappear. The exact $\a$-corrected metric exhibits interesting metric behavior as well, as it becomes approximately anti-de Sitter outside of the region $(r_1,r_2)$. These behaviors must be interpreted with caution, however, because of the unknown higher order $\a$-corrections. 

For additional analysis of the D-term equation, we will work in a different gauge, where instead of choosing the complex gauge transformation $G$ to be Hermitian, we choose it to be of the form 
\begin{equation}
G = UD
\end{equation}
with $U\in \text{SU}(2)$ and $D$ a diagonal matrix with positive eigenvalues. For the monopole-type solution, these matrices are 
\begin{equation}
D=\left( \begin{array}{cc}
e^{u(r)} & 0 \\
0 & e^{-u(r)} \end{array} \right)\:,\;\;\;
U=\left( \begin{array}{cc}
\cos(\theta/2) & e^{i\phi}\sin(\theta/2) \\
-e^{-i\phi}\sin(\theta/2) & \cos(\theta/2) \end{array} \right)\:,
\end{equation}
where $(\theta,\phi)$ are the usual angles of $\mathbb{R}^3$. The function $u(r)$ satisfies the same D-term equation \eqref{eq:monoDterm}. In this gauge, the complex flat connection is 
\begin{equation}
\A =\left( \begin{array}{cc}
-i u'(r) \d r - \sin^2(\theta/2) \d \phi & \tfrac{1}{2} e^{i \phi} e^{-2u(r)}(-i \d\theta + \sin\theta \d\phi) \\
\tfrac{1}{2} e^{-i \phi} e^{2u(r)}(i \d\theta + \sin\theta \d\phi) & i u'(r) \d r + \sin^2(\theta/2) \d \phi \end{array} \right)\:.
\end{equation}
Near sources for the Higgs field, both $u(r)$ and $u'(r)$ blow up, so some components of $\A$ will blow up as well. We can examine the rates at which components of $\cal A$ blow up near the sources at $r=r_1$ and $r=r_2$ to determine the behavior of $\cal A$. In particular, the ratio of $u'(r)$ to $e^{2u(r)}$ determines whether the diagonal or off-diagonal components of $\cal A$ dominate in the limit. The behavior of $u(r)$ is controlled by the D-term equation, so we see that $\a$-corrections may influence the behavior of $\cal A$ near the sources. Numerical results suggest that 
\begin{equation}
\bigm| u'(r)/e^{2u(r)} \bigm| \to c 
\end{equation}
for a constant $c$ that is positive when $u(r)$ satisfies the tree-level or one-loop D-term equation. Thus, all terms in $\cal A$ are of the same order, and the solution is fully non-abelian near the singularities. We may also consider the exact solution to the $\a$-corrected D-term equation, for which there is no singularity at $r_1$ and $r_2 \to \infty$. In this case, we find that $c=0$ for the singularity at $\infty$, so that the off-diagonal components dominate. Furthermore, the lower left component of $\cal A$ dominates the top right one, so that for a fixed $(\theta,\phi)$, the connection sits asymptotically in an abelian subalgebra of ${\mathfrak sl}(2;{\mathbb C})$. However, the dependence of the differentials on $\theta$ and $\phi$ ensure that $\cal A$ sits in a different abelian subalgebra at every point on the celestial sphere. 

The behavior of the Higgs field near sources determines what methods may be used to calculate its spectrum, as will be discussed in the next section. In the present case of the monopole-type solution, its non-abelian behavior near sources prevents us from calculating its spectrum directly, but we may reliably calculate its chiral index, as will be described below. 

\section{Localised Chiral Matter}

Now we will consider the matter spectrum associated to non-abelian solutions to the reduced Hull-Strominger system. In the abelian case, the spectrum is computed via the relative cohomology of $M_3$ with respect to the sources for the Higgs field. For non-abelian solutions, the computation will not always be so straightforward, but in some cases we may apply the same techniques as in the abelian case. 

For our spectrum computations, we will assume a complex flat connection $\cal A$ with trivial holonomy on the three-manifold $M_3$. We write our connection as ${\cal A} = G^{-1}\d G$ with a polar-decomposed gauge transformation $G=\tilde{U}H$, where $\tilde{U}$ is unitary and $H$ is Hermitian with non-negative eigenvalues. We may additionally choose a gauge in which $G=UD$ for a different unitary $U$ and a diagonal $D$.\footnote{Note that we will use the same symbols ${\cal A}$ and $G$ for the connection and $SU(N)$-valued function for different choices of gauge. Because we will fix a certain gauge in each instance, this should not cause ambiguity.} We assume that $\cal A$ satisfies the D-term equation, so that it provides a solution to the reduced Hull-Strominger system. We remark that in this paper we will assume that $G$ is globally defined, but singular at points. The implications of this are discussed further below.  This resembles the common assumption made for abelian Higgs fields $\phi=\d f$ where $f$ is a global but singular function; other generalizations of this abelian setup have been studied in, for example, \cite{Pantev:2009de, Hubner:2020yde}.

The effect of $\a$-corrections on the spectrum computation is only to modify the metric on $M_3$, which may modify the spectrum, but the method itself is independent of which order in $\a$ we consider. Thus, we will not choose a particular order in $\a$ for this section. We leave the case of $\cal A$ with nontrivial monodromies for future work. 

\subsection{Behavior Near Sources}

Our analysis of the matter spectrum is dependent on the asymptotic form of the Higgs field near its sources. We may classify the behavior of such solutions near a source into three cases:
\begin{description}
 \item[$\bullet$ Type 1:] The connection $\cal A$ becomes abelian near the source, meaning that there exists an abelian subalgebra $\mathfrak{h}\subset \mathfrak{g}$ such that the norm of $\cal{A}$ becomes dominated by the components along $\mathfrak{h}$ as one approaches the source. In this case, we may calculate the spectrum using methods analogous to those for an abelian solution.
 \item[$\bullet$ Type 2:] The connection ${\cal A} = G^{-1}\d G$ does not become abelian near the source, but $U$ remains nonsingular, where $U$ is the unitary matrix in $G=UD$. In this case, the tools we develop to compute the spectrum using relative cohomology don't apply, but we may at least compute the chiral index reliably. We can do this by smoothly deforming away $U$ to obtain an abelian Higgs field, which may change the spectrum, but not its index. 
  \item[$\bullet$ Type 3:] The connection $\cal A$ does not become abelian near the source and $U$ becomes singular. In this case, there is currently nothing we can say about the spectrum. 
\end{description}
In this section, we will first consider the computation of the spectrum for Type 1 solutions, and use deformation theory to address the chiral spectrum of Type 2 solutions. The cohomology methods used in this section are a generalization of those introduced for flat solutions in \cite{Pantev:2009de} and \cite{Braun:2018vhk} to certain non-abelian solutions. 

For a Type 1 solution with the Hermitian gauge choice, the flat connection may be written as ${\cal A} =H^{-1}\d H$ in terms of a Hermitian matrix $H$ which approaches the factorized form
\begin{equation}
H\;\;\rightarrow\;\;\left( \begin{array}{cc}
 \tilde D & 0 \\
 0 & \tilde H \end{array} \right)\:,
\end{equation}
where $\tilde D$ is diagonal and gives rise to the abelian part of the connection, which blows up near the source. The other block, $\tilde H$ is Hermitian and gives rise to the rest of the connection, which may remain non-abelian near the source, but has a vanishing contribution to the norm of ${\cal A}$ in this limit. This is the form we will assume a Type 1 solution approaches close to any source. 

Note that for the particular set of equations studied here, i.e. non-abelian solutions to the reduced Hull-Strominger system, we do expect stringy corrections to the equations when we approach the sources. While the flatness condition will be unaffected, as it is derived from an F-term, the D-term stability equation will receive corrections. There is hence a small tubular neighbourhood around any source wherein we do not know what equations we are solving, except that the complex connection remains flat. Since finding the true equations to solve is beyond the scope of this paper, we will instead model the true solution within this neighbourhood by a flat connection of the above type. I.e. an abelian part containing the sources, plus a non-abelian part commuting with the singular part. This can always be done since the boundary of the tubular neighbourhood is a Riemann surface $\Sigma$, and all maps from $\Sigma$ into $SU(N)$ are homotopy equivalent.

\subsection{Matter Field Excitations}

As in the case of abelian solutions, fermions can be represented as poly-forms $\psi_\R\in\Omega^*(M_3,E_\R)$, where $E_\R$ is the vector bundle associated to the representation $\R$ of $SU(N)$ or its complexification, $SL(N;\mathbb{C})$. Chiral fermions correspond to odd forms while anti-chiral fermions are even. They solve the Dirac equation
\begin{equation}
{\cal D}_\A\psi_\R=\d_\A\psi_\R+\d^\dagger_\A\psi_\R=0\:.
\end{equation}
We are interested in solutions that are localised appropriately away from the boundary of the geometry\footnote{We will define more clearly what we mean by appropriately localised away from boundaries below.} or any singularities of the complex flat 
connection. We can then further impose that
\begin{equation}
\d_\A\psi_\R=0\:,\;\;\;\d^\dagger_\A\psi_\R=0\:,
\end{equation}
which follows from a simple integration by parts argument. We may also assume without loss of generality that $\psi_\R$ has a given form degree $p$. These equations are equivalent to
\begin{equation}
\label{eq:LocHarm}
\d\left(G^\alpha \cdot \psi_\R^{\alpha}\right)=0\:,\;\;\;\d\left(({G^\alpha}^\dagger)^{-1} \cdot *\psi^\alpha_\R\right)=0\:,
\end{equation}
in a local patch ${\cal U}_\alpha$, where $G^\alpha \in \Gamma({\cal{U}}^\alpha,SL(N;\mathbb{C}))$ is the local gauge transformation that gives rise to the flat connection $\cal{A}$. Here the dot denotes the action of $G$ on the given representation. (For example, if $\psi_\R$ is in the fundamental representation, it is just matrix multiplication on a vector, whereas if $\R={\rm Ad}(SU(N))$ it is the adjoint action.) The $*$ denotes the three-dimensional Hodge-star. The polar decomposition in a local patch is given as
\begin{equation}
G^\alpha=HU^\alpha\:,\;\;\;({G^\alpha}^{\dagger})^{-1}=H^{-1}U^\alpha\:.
\end{equation}
Hence, solving the equations \eqref{eq:LocHarm} is equivalent to solving the equations
\begin{equation}
\label{eq:NearlyAbHarm1}
\d(H \cdot \tilde\psi_\R)=0\:,\;\;\;\d\left(H^{-1} \cdot *\tilde\psi_\R\right)=0\:,
\end{equation}
for $\tilde\psi_\R=U^\alpha \cdot \psi_\R^{\alpha}$. Note that we have dropped the superscript $\alpha$ as $\tilde\psi_\R$ is a {\it global object}, i.e. $\tilde\psi_\R^\alpha=\tilde\psi_\R^\beta$ on overlaps ${\cal U}_\alpha\cap{\cal U}_\beta$. We will also drop the tilde from here on. 

Our matrix function $H$ is globally defined because we have assumed that $\cal{A}$ has trivial monodromies.  
But even if $H$ were not global, we must still have
\begin{equation}
(H^\alpha)^{-1}\d H^\alpha=(H^\beta)^{-1}\d H^\beta
\end{equation}
on overlaps ${\cal U}_\alpha\cap{\cal U}_\beta$. Using this, and the fact that $H^\alpha$ and $H^\beta$ are positive definite matrices, we see that 
\begin{equation}
\d(H^\alpha \cdot \psi_\R)=0\;\;\;\Leftrightarrow\;\;\;\d(H^\beta \cdot \psi_\R)=0
\end{equation}
on overlaps. With this in mind we may as well take $H^\alpha$ and try to extend it to a full global solution $H$. The obstructions for doing so will be the monodromies of $\cal{A}$, which can be trivialized by removing a submanifold of positive codimension, analogous to a branch cut. Such an operation will modify the boundary conditions for the Higgs field, which affects the spectrum. 

Now we will go to a gauge where $G=UD$, where $U$ is unitary and $D$ is diagonal. Again, because $\cal{A}$ has trivial monodromies, both $U$ and $D$ are globally defined. In this gauge we have
\begin{equation}
\label{eq:NearlyAbHarm}
\d\left((UD) \cdot \psi_\R\right)=0\:,\;\;\;\d\left((UD^{-1}) \cdot *\psi_\R\right)=0\:,
\end{equation}
or equivalently
\begin{equation}
\label{eq:NearlyAbHarm2}
\d_A(D \cdot \psi_\R)=0\:,\;\;\;\d_A(D^{-1} \cdot *\psi_\R)=0\:,
\end{equation}
where $A=U^{-1}\d U$ is a flat connection. 
If $D$ is regular on $M_3$, then solutions to the set of equations \eqref{eq:NearlyAbHarm1} are counted by the cohomology
\begin{equation}
H^*_{\d_A}(M_3;\R)\cong H^*_{\d}(M_3;\R)
\end{equation}
where the isomorphism is due to the fact that $A$ is a globally trivial flat connection. However, as in the abelian case \cite{Pantev:2009de}, if the eigenvalues of $D$ blows up, we need to restrict to solutions $\psi_{\cal R}$ with appropriate vanishing properties at those singular loci. This means that we should compute a relative cohomology. 

\subsection{The Relative Cohomology}
We are now in a position to define the relative cohomology in question. Let us rewrite the closure equation in analogy with the abelian case as 
\begin{equation}
\d(D\tilde\psi)=0\:,
\end{equation}
where we have defined
\begin{equation}
\tilde\psi=(D^{-1}UD)\cdot\psi\:,
\end{equation}
and we have dropped the $\R$-label on the fields. Note that because $U$ factorizes at the singularities, and in particular becomes the identity matrix for the eigenvalues corresponding to sources, the $SL(N;\mathbb{C})$-valued invertible matrix $D^{-1}UD$ is regular over the three-manifold. Indeed, in the region of singularities $U$ becomes block-diagonal and the identity matrix for the given eigenvalues of $D$ which blow up or vanish. As in the abelian case, we will require the component $\tilde\psi_i$ or equivalently $\psi_i$ to vanish where the corresponding eigenvalue $\lambda_i$ blows up. 

The co-closure equation reads
\begin{equation}
\d(D^{-1}*\tilde H \cdot \tilde\psi)=0\:,
\end{equation}
where we have defined the Hermitian metric on the bundle
\begin{equation}
\tilde H=DUD^{-2}U^\dagger D\:.
\end{equation}
We note that by the above reasoning, the metric $\tilde H$ is also regular over $M_3$. In particular, it approaches a block diagonal form near the sources where it becomes the identity matrix for the given eigenvalues of $D$ which blow up or vanish. Again, as in the abelian case we require the component $*\tilde\psi_i$ or equivalently $*\psi_i$ to vanish where the corresponding eigenvalue $\lambda_i$ goes to zero.

We now come to defining the relative cohomology. Let us first comment on what we take to be the domain of the component $\tilde\psi_i$ of $\tilde\psi$ corresponding to the $i$th eigenvalue of $D$. As in the abelian case, we will take this to be $M_3^i=M_3\setminus\Delta_i^+\cup\Delta_i^-$, where $\Delta_i^+$ and $\Delta_i^-$ are small tubular neighbourhoods of the positive and negative sources corresponding to $\lambda_i\rightarrow\infty$ and $\lambda_i\rightarrow0$ respectively. We can then define the inner-product
\begin{equation}
(\psi^1,\psi^2)=\sum_i\int_{M^i_3}\bar\psi_i^1\wedge* (\tilde H\cdot\psi^2)_i\:,
\end{equation}
where the bar denotes complex conjugation. We will denote $\partial\Delta_i^\pm=\Sigma_i^\pm$. Note that as we approach these boundaries, $(\tilde H \cdot \psi^2)_i\rightarrow\psi^2_i$.

Note that the harmonic types of $\R$-valued $k$-forms $\tilde\psi$ we are considering (where $\tilde\psi$ vanishes when restricted to $\Sigma^+$ and $*\tilde\psi$ when restricted to $\Sigma^-$)\footnote{Note that a form vanishing when restricted to a sub-manifold $\Sigma$ does not imply that its  Hodge-dual will vanish as well, because normal components of the form might still be non-zero.} form part of a Hodge-type decomposition of forms
\begin{equation}
\{\tilde\psi\}\oplus\{\d_D\beta\}\oplus\{\d_D^{\tilde\dagger}\gamma\}\:,
\end{equation}
with respect to the above inner-product. Here $\beta$ is an $\R$-valued $(k-1)$-form and $\gamma$ is an $\R$-valued $(k+1)$-form. However, this does not span all the allowed forms, as restrictions are put on $\tilde\psi$, $\beta$, and $\gamma$ at the boundaries. Here $\d_D=D^{-1}\circ\d\circ D$, and $\d_D^{\tilde\dagger}=\tilde H^{-1}D\circ\d^\dagger\circ D^{-1}\tilde H$ is the adjoint of $\d_D$ with respect to the above inner product. In addition to the above restrictions on $\tilde\psi$, we also restrict $\beta_i$ to vanish at the positively charged boundaries $\Sigma_i^+$, while $*\gamma_i$ vanishes at the negatively charged boundaries $\Sigma_i^-$. We can confirm that the individual components are orthogonal with respect to the inner product. For example
\begin{align}
(\tilde\psi,\d_D^{\tilde\dagger}\gamma)= \sum_i\int_{M^i_3}\bar{\tilde\psi}_i\wedge*(D\d^\dagger D^{-1}\tilde H\cdot\gamma)_i\notag =- \sum_i\int_{\Sigma_i^+\cup\Sigma_i^-}\bar{\tilde\psi}_i\wedge*(\tilde H\cdot\gamma)_i\:,
\end{align}
where we have integrated by parts and used that $\tilde\psi$ is harmonic and so $\d_D$-closed. Because the Hermitian metric $\tilde H$ becomes the identity on the boundaries, we end up with
\begin{equation}
(\tilde\psi,\d_D^{\tilde\dagger}\gamma)=- \sum_i\int_{\Sigma_i^+\cup\Sigma_i^-}\bar{\tilde\psi}_i\wedge* \gamma_i=0\:,
\end{equation}
since $\tilde\psi_i$ vanishes at $\Sigma^+_i$ while $* \gamma_i$ vanishes at $\Sigma^-_i$. It can be checked that the other terms in the Hodge decomposition are similarly orthogonal. 

We then claim that the $\tilde\psi$ that give rise to stable complex flat connections are in one to one correspondence with the relative cohomology classes of $\d_D$, which acts on forms that vanish on the positive boundaries exactly as in the abelian case. Indeed, consider a $\d_D$-closed form $\alpha$ which vanishes at the positive boundaries and is orthogonal with respect to the above inner product to the set $\{\d_D\beta\}$ in the above Hodge decomposition. We require
\begin{equation}
0=(\d_D\beta,\alpha)=(\beta,\d_D^{\tilde\dagger}\alpha)+ \sum_i\int_{\Sigma_i^-}\bar\beta_i\wedge*\alpha_i\:.
\end{equation}
If this is to vanish for all $\beta$ (which vanish appropriately at positive boundaries), we see that we need to require $\d_D^{\tilde\dagger}\alpha=0$ in addition to $*\alpha_i$ vanishing at the negative boundaries $\Sigma_i^-$. Hence, $\alpha$ is harmonic with respect to the above definition. The one to one correspondence between harmonic forms and cohomology classes follows. 

So we see that to find the spectrum we proceed just as in the abelian case. We find the chiral and anti-chiral modes are counted using the eigenvalues of $\log(D)$ in the appropriate representation as Morse functions when computing the relative cohomology. We may hence compute the number of chiral zero modes $N_{\chi_\R}$ and anti-chiral zero-modes $N_{\bar\chi_\R}$ in this representation as
\begin{align}
N_{\chi_\R}&=\sum_{i=1}^{n_\R} h^1(M^i_3,\Sigma_i^+)\\
N_{\bar\chi_\R}&=\sum_{i=1}^{n_\R} h^2(M^i_3,\Sigma_i^+)\:,
\end{align}
where we also note that as in the abelian case the zeroth and third order cohomologies vanish. Indeed, for the zeroth order cohomology, for instance, we are again looking for constant functions which vanish at the boundaries of the charged regions, or global harmonic functions if the Morse functions for the given representation are regular. This of course can only happen if the function vanishes. Further details of how to compute these cohomologies for a given source configuration are given in \cite{Pantev:2009de, Braun:2018vhk}. 

The index counting the net chirality is then
\begin{equation}
\label{eq:Index}
{\rm Index}({\cal D}_\R)=N_{\chi_\R}-N_{\bar\chi_\R}\:,
\end{equation}
for the Dirac operator ${\cal D}_\R$ in the representation $\cal R$. We finish this section by remarking that even if the above mentioned factorisation of the complex flat connection into an abelian and a non-abelian part near the charged sources should not happen for a particular solution, we still expect the net chiral index to be counted by equation \eqref{eq:Index}. Indeed, as we will see below, given a complex flat connection in terms of an $SL(N;\mathbb{C})$-valued matrix $G=UD$, assuming a regular $U$-matrix so we are not infinite far away from a diagonal $G$ in deformation space, and keeping the behaviour of $D$ near the sources fixed, we can always smoothly deform $G$ to be diagonal. On general grounds we expect the index to be topological, and hence insensitive to such deformations. In the end we expect the index to compute the number of massless modes in $\R$, and we expand on this in section \ref{sec:examples}.

\subsubsection*{A Comment on Yukawa Couplings}
Methods for computing Yukawa couplings have been developed in the case of stable abelian complex flat connections using the gradient flow trajectories of the Morse functions \cite{Pantev:2009de, Braun:2018vhk}. These Yukawa couplings are given by multi-linear maps from the cohomologies computing the chiral spectrum into $\mathbb{C}$. For example, the third order couplings are of the form
\begin{equation}
{\rm Yuk}({\R_1,\R_2,\R_3}): H^1(M_3;\R_1)\times H^1(M_3;\R_2)\times H^1(M_3;\R_3)\rightarrow \mathbb{C}\:,
\end{equation}
where there must be at least one non-trivial singlet in the tensor product $\R_1\otimes\R_2\otimes \R_3$. The Yukawa couplings in the non-abelian case are given by similar multi-linear maps. Using the isomorphism between the cohomologies, we can hence map a non-abelian Yukawa to an abelian Yukawa, where we can use the gradient flow method to compute the coupling. 

\subsection{Deforming Solutions}
Having seen in the previous section that abelian and non-abelian solutions are closely related in the way the chiral spectrum is computed, one may wonder if one can deform a given non-abelian background to an abelian one. Equivalently, are the non-abelian solutions in question simply deformations of abelian ones? We will argue here in the affirmative of this, at least for non-abelian solutions that are well-behaved enough. As before, assume an $SU(N)$-valued function $G=UH$ in polar decomposed form, where $U$ is unitary and $H$ is a positive matrix. We may write $H=\tilde{U}^{-1}D\tilde{U}$ for a unitary matrix $\tilde{U}$ and a diagonal matrix $D$ of positive eigenvalues. Now let $U'=U\tilde{U}^{-1}$, so that $G=U'DU$. Our 'well-behaved' assumption is then that $U'$ is non-singular, which corresponds to the Type 2 classification described at the beginning of this section. 
 
To turn a non-abelian complex flat connection into an abelian one is then equivalent to turning off $U'$ (as we may eliminate the other unitary matrix, $U$, by a gauge transformation). We note that a given $U'$ corresponds to a map from $M_3$ into $SU(N)$ and hence has a representative homotopy class in $[M_3,SU(N)]$. For $M_3=S_3$ this is $\pi_3(SU(N))=\mathbb{Z}$. If the representative of $U'$ is the trivial class, we can always turn off $U'$ by deformation. If the class of $U'$ is non-trivial, we can deform $U'$ to be the identity matrix inside of a tubular neighborhood of the singularities of the eigenvalues of $D$. Indeed, we assume the singularities have co-dimension at least one in $M_3$, and $U'$ can always be trivialised on a graph or Riemann surface. Simultaneously, we deform $D$ to be the identity matrix outside of the tubular neighborhood. Thus, at every point, either $U'$ or $D$ is the identity matrix, so that the deformed matrices commute. We may then write $G=DU'U$, and the unitary factor $U'U$ may be removed by a global gauge transformation, leaving $G$ as diagonal, and thus giving rise to an abelian connection. 

In the deformations involved in the above steps, the chiral spectrum may be modified, but the chiral index should remain the same, because it is invariant under smooth deformation. Thus, in the case of Type 2 non-abelian complex connections, we expect that the chiral index of the spectrum may be reliably computed by an abelian deformation.  
 
The above describes a way of deforming a given non-abelian flat connection to an abelian one, or equivalently a deformation of an abelian solution to a non-abelian one. This is a solution to the Maurer-Cartan equation 
\begin{equation}
\label{eq:MCeq}
\d_\A\alpha+\,\alpha\wedge\alpha=0\:,
\end{equation}
where $\alpha$ is the deformation of the connection. We keep the part of the connection which blows up at a source fixed at this source under such deformations. Recall that this part is assumed to become abelian, commuting with the rest of the connection. For the deformed connection to satisfy \eqref{eq:CoclosureZero}, we must also impose that the deformation of the co-closure condition holds. Before we consider this equation, we note that $\A$ should be thought of as a  {\it holomorphic function} on the complex parameter space of complex connections. Indeed, as shown in \cite{Acharya:2002kv}, reduction of the six-dimensional holomorphic Chern-Simons functional gives the superpotential 
\begin{equation}
W(\A)=\int_{M_3}\tr\left(\A\wedge\d\A+\,\tfrac{2}{3}\A\wedge\A\wedge\A\right)\:.
\end{equation}
Hence, when considering such a holomorphic deformation $\Delta$, we should set $\Delta\A=\alpha$ and $\Delta\A^\dagger=0$. Imposing this on the co-closure condition for such deformations, we find
\begin{equation}
\label{eq:DefHolCoClosure}
(\d_\A)^\dagger\alpha=0\:.
\end{equation}
This equation imposes that the $\d_{\A}$-exact part of $\alpha$ in the Hodge decomposition 
\begin{equation}
\alpha=\alpha_h+\d_{\A}\beta+\d_{\A}^\dagger\gamma
\end{equation}
vanishes. Given a solution to the Maurer-Cartan equation \eqref{eq:MCeq}, we can always do a Laurent series type epansion
\begin{equation}
\alpha(Z)=Z^A\alpha_A+\tfrac12Z^AZ^B\alpha_{AB}+..\:,
\end{equation}
where $Z^A$ denote holomorphic coordinates on the moduli space. If $\alpha$ then contains a $\d_A$-exact part, it can be checked that this can be removed by redefining $\alpha$ order by order in this expansion, thus also solving the co-closure equation. Holomorphy and the Laurent theorem then guarantees that the resulting series can be re-summed. 

We see that when considering a deformation in a holomorpic direction $\Delta$, the first order deformations $\alpha_A$ are harmonic and hence counted by the relative cohomologies
\begin{equation}
N_{Ad(N)}=\sum_{i=1}^{n_{Ad(N)}}h^1(M_3^i;f^i_{{\rm Ad}(N)})\:,
\end{equation}
where $f^i_{{\rm Ad}(N)}$ are the Morse functions of the adjoint representation $Ad(N)$ of $SU(N)$. In this paper we have restricted to flat backgrounds of the form $\A=G^{-1}\d G$ where $G$ is global, but with singularities, as a direct generalisation of abelian solutions. We want to preserve this ansatz under deformations, so in general not all of the above deformations will be considered. To find the relevant deformations, consider part of the long exact sequence used to compute $H^i(M_3^i,\Sigma^+_i)$
\begin{equation}
\label{eq:LES1}
...\rightarrow H^q(M^i_3,\Sigma^+_i)\xrightarrow{i^*_p} H^q(M^i_3)\xrightarrow{p^*_p} H^q_{\d}(\Sigma^+_i)\xrightarrow{\alpha_p} H^{q+1}(M^i_3,\Sigma^+_i)\rightarrow...\:.
\end{equation}
By the long exact sequence we have
\begin{equation}
H^1(M^i_3,\Sigma^+_i)\cong {\rm Im}(\alpha_0)\oplus {\rm Im}(i^*_1)\:.
\end{equation}
the deformations which preserve the global triviality of the complex flat connection correspond to the modes in the image of the connecting homomorphism $\alpha_0$ in the above direct sum. Assuming each $M_3^i$ are connected, there are precisely $n^+_i-1$ such modes, where $n^+_i$ denotes the number of connected positively charged regions in $M_3^i$, i.e. the number of components where the given Morse function $f^i_{Ad(N)}$ blows up. As explained above, these directions are necessarily unobstructed.  

Writing the deformation $\alpha$ as an $N\times N$-matrix, there is a Morse function corresponding to each element $\alpha_{pq}$, which takes the form
\begin{equation}
f_{pq}=f_p-f_q\:,
\end{equation}
where $f_p=\log(\lambda_p)$ is the Morse function of the $p$th eigenvalue of $\log(D)$ as an $N\times N$ matrix. We are hence counting the number of positive sources for these Morse functions. Note in particular that starting with an abelian solution, the Morse functions for the diagonal directions that keep the solution abelian vanish, and we need to turn on off-diagonal non-abelian directions in order to deform the solution non-trivially.

Had we instead considered real deformations $\Delta+\bar\Delta$ of the complex flat connection, the zeroth order co-closure equation becomes
\begin{equation}
\label{eq:DefCoClosure}
\left(\d_\A^\dagger\alpha-\d^\dagger_{\A^\dagger}\alpha^\dagger\right) - [\alpha^m,\alpha^\dagger_m]=0\:.
\end{equation}
At first glance, this equation may put potential obstructions on the deformations $\alpha$. However, as we argued above the zeroth order Morse cohomologies vanish. Using this fact this equation may also be solved order by order by the usual methods of perturbative deformation theory.

\section{Matter Spectrum Examples}
\label{sec:examples}
Before we move to discuss examples, let us first see why in the end it is the index of the Dirac operator ${\cal D}_\A$ which counts the number of massless states in the final low-energy theory. In a heterotic compactification on a six-dimensional $SU(3)$ structure manifold $X$, the zero-modes in a representation $\R$ are counted by the cohomology $H^{(0,1)}_{\bp_A}(X;\R)$. Reality considerations of the decomposition of the gauge group implies that for $(0,1)$ modes in $\R$ we also expect to look for $(0,1)$ modes in $\bar\R$, i.e. counted by $H^{(0,1)}_{\bp_A}(X;\bar\R)$. Such modes can couple in the super-potential. For example, a fundamental mode $a^c$ can couple with an anti-fundamental mode $b_d$ via an adjoint mode ${\alpha^d}_c\in H^{(0,1)}_{\bp_A}(X;\End(V))$ in a Yukawa coupling of the form
\begin{equation}
{\rm Yuk}(a,b,\alpha)=\int_X\,a^c\wedge b_d\,\wedge{\alpha^d}_c\wedge\Omega\:,
\end{equation}
where $\Omega$ is the holomorphic top-form on $X$. Generically, such couplings are expected to remove modes in $\R$ and $\bar\R$ in pairs, such that the true massless spectrum is counted by the index of $\bp_A$ in the given representation. 

Lets see how this works in the reduced three-dimensional setting. In the three-dimensional theory a $(0,1)$-mode in the representation $\R$ reduces to a one-form mode on the three-manifold $M_3$, i.e. an element of the cohomology $H^1(M_3;\R)$. Note then that in six dimensions, taking the Hodge dual of a $\bp_A$-harmonic $(0,1)$-modes in $\bar\R$ correspond to a $\p_A$-harmonic $\bar\R$-valued $(2,3)$-form which also corresponds to a  $\p_A$-harmonic $\bar\R$-valued $(2,0)$-form. Complex conjugation then gives a $\bp_A$-harmonic $\R$-valued $(0,2)$-form. In the reduction these give rise to $\R$-valued two-forms modes on the three-dimensional space,  i.e. elements of the cohomology $H^2(M_3;\R)$. We therefore expect couplings between such one and two-form modes, and the true massless spectrum is computed by the index of $\d_{\A}$ in the $\R$-representation.

\subsection{Example: Breaking patterns of $SU(4)$}
To get a feel for how this goes, let us consider breaking $SU(4)$ to $SU(3)$ by turning on an abelian Higgs field corresponding to an $SU(4)$ generator of the form $\log(D)={\rm diag}(f,f,f,-3f)$, and consider the spectrum of ${\rm Ad}(SU(4))$. A similar example breaking $SU(6)$ to $SU(5)$ is given in \cite{Pantev:2009de}. We have
\begin{equation}
{\rm Ad}(SU(4))={\rm Ad}(SU(3))+{\bf 3}_{1} + \bar{\bf 3}_{-1} + {\bf 1}\:.
\end{equation}
The subscripts denote the charges under the (broken) $U(1)$. For such a configuration, one finds that the adjoint action of $SU(4)$ gives $f$ as the Morse function counting modes in the fundamental representation $\bf 3$, while $-f$ counts the  modes in the anti-fundamental one $\bf {\bar 3}$.

We can instead consider breaking $SU(4)\rightarrow U(1)$ by turning on a non-trivial $SU(3)$ within $SU(4)$, e.g. by choosing a non-abelian configuration with $\log(D)={\rm diag}(f_1,f_2,-f_1-f_2,0)$. Chiral and anti-chiral modes of positive charge (transforming in the fundamental of the $SU(3)$) are then computed by the Morse cohomologes of $f_1$, $f_2$ and $f_3=-f_1-f_2$, while $-f_i$'s compute negatively charged modes. For example, one could imagine a model on $S^3$ where both $f_1$ has a single positive point source and a single negative point source, while $f_2$ has a single positive source overlapping the one of $f_1$ and a single negative point source. Consider positively charged matter. A straight forward computation in relative cohomology shows that we get
\begin{equation}
h^1(M_3;f_3)=1\:,
\end{equation}
while all other $h^{1/2}(M_3;f_i)$ vanish. Hence, such a model would give a single chiral mode of positive charge, and the chiral index in this representation is one. 

\subsection{Example: Monopole-type Solution}
Let's consider the monopole-type solution on ${\mathbb{R}}^{3}$ described in section \ref{sec:ExamplesFlat} above. Let's consider the zeroth and first order (non-exact) solutions for now, where the eigenvalues have finite radius singularities. For this case we have
\begin{equation}
D=\left( \begin{array}{cc}
e^u & 0 \\
0 & e^{-u} \end{array} \right)\:,\;\;\;
U=\left( \begin{array}{cc}
\cos(\theta/2) & e^{i\phi}\sin(\theta/2) \\
-e^{-i\phi}\sin(\theta/2) & \cos(\theta/2) \end{array} \right)\:,
\end{equation}
where $(\theta,\phi)$ are the usual angles of $\mathbb{R}^3$ and $u$ satisfies the equation \eqref{eq:RadExLapl} in the tree-level case or the $O(\a)$ part of \eqref{eq:monoDterm} in the $\a$-corrected case. Set $\lambda_1=e^u$ and $\lambda_2=e^{-u}$. Let's consider computing the spectrum of the fundamental representation in $SU(2)$. Note that we then need to consider the modes of both $f_1=u$ and $f_2=-u$. Let's consider some examples of configurations of $u$ for this setup. But before we do, we note that the zeroth order solution in $\a$ for this ansatz does not satisfy the assumption that the solution becomes abelian near the sources (though the exact $\a$-corrected solution does satisfy this criteria). The individual cohomology computations at zeroth order might hence be less trustworthy. Note however that as the matrix $U$ is regular, we still expect the index computation to be reliable as we can smoothly deform $U$ to the identity. 

To avoid issues concerning the $U$ being ill-defined at $r=0$, let's consider an example where $u$ tends to negative infinity at a small finite $r$ and blows up at a larger $r$. The space is now a three-dimensional annulus, that is, an open ball with a hole at the origin. It is easy to compute both
\begin{equation}
h^1(M_3;f_{1/2})=0\:,\;\;\;h^2(M_3;f_{1/2})=0\:,
\end{equation}
and so this geometry has no chiral or anti-chiral modes. Let us also consider a $u$ that blows up at a small $r$ and a large $r$. The space is again a three-dimensional annulus. There are again no zero modes for degree zero and three, but we now find for $f_1$
\begin{equation}
h^1(M_3;f_1)=1\:,\;\;\;h^2(M_3;f_1)=0\:.
\end{equation}
so we have a chiral zero mode. From a relative cohomology perspective, this mode corresponds to exact forms $\d f$ where $f$ approaches different constant values on the boundaries $\partial^+M_3$. However, if we consider $f_2=-u$ we find
\begin{equation}
h^1(M_3;f_2)=0\:,\;\;\;h^2(M_3;f_2)=1\:,
\end{equation}
so the net chirality or index of the solution is zero. 

Next, we consider a solution of $f_1$ that vanishes at infinity and blows up at finite radius. In order to avoid complications regarding harmonic modes on non-compact geometries, we assume that the solution can be embedded in a large three-sphere, with a flat metric where the solution is non-trivial. Hence our space is $S^3$ minus an open ball. Again, we find
\begin{equation}
h^1(M_3;f_{1/2})=0\:,\;\;\;h^2(M_3;f_{1/2})=0\:,
\end{equation}
and so the chiral index of the spectrum is trivial for this solution as well. 

\subsection{Chiral index of $SU(2)$ representations}
The vanishing of the chiral index for $SU(2)$ representations is actually much more general. Indeed, consider some $SU(2)$ solution which we have deformed to an abelian solution by the procedure described above. Then given a normalisable chiral mode $\psi_0\in\Omega^1(M_3,{\bf 2})$ of the form
\begin{equation}
\psi_0=\left( \begin{array}{c}
\psi_a \\
\psi_b \end{array} \right)\:,
\end{equation}
solving the Dirac equation, it is clear that the anti-chiral mode
\begin{equation}
\tilde \psi_0=\left( \begin{array}{c}
*\psi_b \\
*\psi_a \end{array} \right)
\end{equation}
will  solve the Dirac equation as well. Hence the Chiral index of the fundamental representation of $SU(2)$ solutions vanishes.  It can be checked that this is also the case for the adjoint representation $\bf 3$, and is thus also true for the singlet as
\begin{equation}
{\bf 2}\otimes {\bf 2}= {\bf 1}+ {\bf 3}\:.
\end{equation}
Inductively, it is easy to convince oneself that the chiral index will vanish for all higher irreducible representations as well, by taking higher tensor products with the fundamental representation. Hence it does not appear that $SU(2)$ solutions on $\mathbb{R}^3$ or $S^3$ support a non-trivial chiral index in any representation.  

One can imagine embedding $SU(2)$ into a larger group, and in this way achieve a spectrum of non-vanishing chiral index. For example, let's assume that in the example above, the fundamental representation is also charged with respect to a $U(1)$ whose Morse function is $2u$.\footnote{This Morse-function will generically obey a different co-closure equation, but we take it to be $2u$ for illustration purposes.} The solution where $g$ tends to infinity at a small and a large $r$ would then have two chiral modes and zero anti-chiral modes in the representation ${\bf 2}_1$.

\subsection{Example: Non-abelian T-brane solution of Barbosa et al.}
Of course, it may happen that the complex flat connection cannot be written as $\A=G^{-1}\d G$ for a global (but singular) $G$. This is the case when $\A$ has monodromies of various sorts. An example of this kind turns out to be the non-Abelian local solution of \cite{Barbosa:2019bgh}. Let us briefly discuss this example now. 
 
The non-abelian explicit solution constructed in \cite{Barbosa:2019bgh} is an $SU(3)$ example on $\mathbb{R}^3$ with coordinates $(x,y,t)$. The authors consider the decomposition $SU(3)\rightarrow SU(2)\times U(1)$ and the complex connection ${\cal A} = A^x + i A^\phi$ with
\begin{align}
A^x &= \left( \begin{array}{ccc}
\tfrac{1}{2} \left( \partial_z f \, \d z - \partial_{\bar z} f \, \d {\bar z} \right) & 0 & 0 \\
0 & -\tfrac{1}{2} \left( \partial_z f \, \d z - \partial_{\bar z} f \, \d {\bar z} \right) & 0 \\
0 & 0 & 0
\end{array} \right) \\
A^\phi &= \left( \begin{array}{ccc}
\tfrac{i}{3}\d h & -v {\bar z} e^{-f(z,{\bar z})} \d {\bar z} + \varepsilon \, e^{f(z,{\bar z})} \d z & 0 \\
v z e^{-f(z,{\bar z})} \d z - \varepsilon \, e^{f(z,{\bar z})} \d {\bar z} & \tfrac{i}{3}\d h & 0 \\
0 & 0 & -\tfrac{2 i}{3} \d h
\end{array} \right) \, ,
\end{align}
where $\varepsilon$ and $v$ are real constants, $h(z,{\bar z},t)$ is the function  
\begin{equation}
h = \tfrac{\kappa}{8} (z + {\bar z})^2 - \tfrac{\kappa}{2} t^2
\end{equation}
for a real constant $\kappa$, and the D-term condition demands that $f(z,{\bar z})$ satisfies the differential equation
\begin{equation}
\tfrac{1}{4}\left( f_{rr} +\tfrac{1}{r} f_r \right) = \varepsilon^2 \, e^{2f} - v^2 r^2 e^{-2f} \, ,
\end{equation}
where $r = \left| z \right|$. In the case where $\varepsilon \neq 0$, this equation may be transformed into a Painlev\'{e} III differential equation, while in the $\varepsilon = 0$ case it becomes a modified Liouville equation. We may attempt to apply our methods to solutions of this form, but we will see that we run into obstacle for computing the spectrum for both choices of $\varepsilon$.

\subsection*{\bf Case 1: $\varepsilon \neq 0$}

To analyze this solution, we want to examine limiting forms of the connection near the sources. In the case of $\varepsilon \neq 0$, the only source is at infinity, so we consider the solution at large distances in the $(x,y)$-plane and at large $t$, with the intention of studying the asymptotic behavior of the Morse functions to determine a charge distribution at infinity. If we take the limit where $r=\vert z\vert$ and $t$ are large, then the asymptotic behavior of the Painlev\'{e} III transcendental reveals the approximate form
\begin{equation}
{\cal A} \to 
\left( \begin{array}{ccc}
-\tfrac{1}{3}\d h +\tfrac{\d z}{8z} - \tfrac{\d {\bar z}}{8{\bar z}}  & -ipr^{-1/2}{\bar z}\d {\bar z} + ipr^{1/2} \d z  & 0 \\
ipr^{-1/2}z \d z - ipr^{1/2} \d {\bar z} & -\tfrac{1}{3}\d h - \tfrac{\d z}{8z} + \tfrac{\d {\bar z}}{8{\bar z}} & 0 \\
0 & 0 & \tfrac{2}{3}\d h
\end{array} \right)\:,
\end{equation}
where $p = \sqrt{\epsilon v}$. With this asymptotic form of ${\cal A}$, the equation $\d G = G\cal{A}$ can be solved explicitly, and we find the asymptotic gauge transformation 
\begin{equation}
G= \tfrac{i+\sqrt{3}}{2}
\left( \begin{array}{ccc}
e^{i\theta/4 - h(z,t)/3} \cosh s  & -e^{-i\theta/4 - h(z,t)/3} \sinh s  & 0 \\
i e^{i\theta/4 - h(z,t)/3} \sinh s & -i e^{-i\theta/4 - h(z,t)/3} \cosh s & 0 \\
0 & 0 & e^{2h(z,t)/3}
\end{array} \right)\:,
\end{equation}
where 
\begin{equation}
s = \tfrac{4}{3} p r^{3/2} \sin(3\theta/2) \: .
\end{equation} 
As discussed above, in order to compute the spectrum as if the solution were abelian we need to check that the solution becomes asymptotically abelian at the sources. By examining the asymptotic form of ${\cal A}$, we see that it will be dominated by the diagonal, and thus asymptotically abelian, as long as we approach infinity in a direction where $\cos \theta \neq 0$. The directions in which the source does not become asymptotically abelian are of measure zero on the celestial sphere, but it is unclear if this fully justifies the use of abelian methods.

However, we also see that despite the fact that $\cal A$ is a flat connection on $\mathbb{R}^3$, our $G$ is not single-valued due to the fractional dependence on $\theta$, and instead should be defined on a four-sheeted cover. This means that we cannot work with a globally defined $G$, so that we cannot apply our method for computing spectra or the chiral index. It would be interesting to generalise the methods we have proposed to this setting, and we leave this for future work. 

\subsection*{\bf Case 2: $\varepsilon = 0$}

We may also consider the $\varepsilon=0$ case, where the D-term equation for $u(r)$ becomes a modified Liouville equation whose solution has a singularity at $r_0 = 1/\sqrt{v}$. This means that the domain of the solution is all of $\mathbb{R}^3$ exterior to a cylinder at $r=r_0$ that extends infinitely in the positive and negative $t$ directions. Our method for computing the chiral matter is hampered in this case by the presence of a nontrivial monodromy around any loop enclosing the cylinder. The Wilson loop around a circular loop $L_r$ at fixed $r$ may be computed as
\begin{equation}
W({\cal A},L_r) = \text{Tr}\exp \left( { - {\int_0^{2\pi}}\d \theta{\cal A}_\theta} \right) = 1+2\cos\tfrac{4\pi v^2 r^4}{1-v^2r^4} \ ,
\end{equation}
where the path ordering is trivial because the quantity in the exponent lies in an abelian subalgebra. In particular, this holonomy comes from the gauge field, while the Higgs field has trivial holonomy. As $r$ approaches $r_0$, the value of the Wilson loop oscillates rapidly, so that the gauge field has a singularity at $r_0$ as does the Higgs field. 

A consequence of nontrivial monodromy is that we cannot find a global solution $G$ to the equation ${\cal A} = G^{-1}\d G$, which again means that the direct computation of the chiral spectrum would require more sophisticated techniques than those presented in the previous section.

\section{Conclusions and Outlook}
In this paper we have considered heterotic string compactifications on $SU(3)$-structure manifolds and their reduction to the equations for a stable complex flat connection on a three-dimensional submanifold. We saw that upon reduction of the heterotic equations, the complex connection remains flat, but the D-term co-closure condition gets corrected even at first order in $\a$ due to torsional effects of the 6D Hull-Strominger geometry and the non-trivial heterotic Bianchi identity. In this paper we have studied local solutions on $\mathbb{R}^3$ for the $\a$-corrected system, including both abelian and non-abelian examples of bundles and their back-reaction on the geometry. These solutions constitute new local $T^3$-invariant solutions to the Hull-Strominger system. It would be interesting to consider the reduced Hull-Strominger system also for more generic compact three-manifolds, and to investigate both the physical and mathematical/topological implications of these corrections. 

We also introduced a way of computing the spectrum (or at least the index) of a particular set of non-abelian solutions with a non-flat gauge field, with the caveat that the complex flat connection $\A=G^{-1}\d G$ is given by a {\it global but singular matrix G}, similar to the common assumption made for abelian Higgs fields $\phi=\d f$ where $f$ is a global but singular function.\footnote{It would be interesting to see how our methods may generalise to the case when $\A$ has non-trivial monodromies, and we leave this to future work.} Assuming an appropriate behaviour of $G$ near singularities (Type 1 or 2 in the classification of section 4.1), the index computation resembles that of abelian Higgs fields. We find that a monopole-type nonabelian solution has vanishing chiral index, while the non-abelian local example of \cite{Barbosa:2019bgh} does not have the correct behavior for our current methods to apply. 

Another avenue to explore is to ask what these $\a$-corrections correspond to on the M-theory side, following the M-theory/heterotic duality. Given that $\a$-corrections correspond to higher curvature corrections from a supergravity point of view, it is conceivable that they correspond to higher curvature corrections on the M-theory side as well, possibly with an incorporation of four-form $G$-flux. Since $\a$-corrections are vital for understanding properties of heterotic compactifications such as the moduli problem, Yukawa couplings, and moduli metric, both in the three-dimensional reduced system and also the upstairs Hull-Strominger geometry \cite{Anderson:2010mh, Anderson:2014xha, delaOssa:2014cia, Garcia-Fernandez:2015hja, delaOssa:2015maa, Candelas:2016usb, McOrist:2016cfl, Ashmore:2018ybe}, a better understanding of these corrections might therefore lead to better insight into such issues on the M-theory side as well. 

Recently, much progress has been made in the study of the heterotic moduli problem \cite{Anderson:2010mh, Anderson:2014xha, delaOssa:2014cia, Garcia-Fernandez:2015hja, Garcia-Fernandez:2018emx, Ashmore:2018ybe, Garcia-Fernandez:2018ypt, Garcia-Fernandez:2020awc}. It would be interesting to reduce the corresponding moduli structures to the three-dimensional setting as well, and study the resulting equations generalising that of the moduli problem of a complex flat connection. Indeed, the geometric structures and moduli problem in particular are expected to retain much of the important physical and mathematical properties of the upstairs geometry, hence the reason for doing such a reduction in the first place. Moreover, the three-dimensional system has the advantage of being much more explicit, which gives hope for a more hands on approach for understanding the geometric structures and moduli, particularly in terms of explicit solutions. It was recently discovered that the moduli problem of the Hull-Strominger system is governed by an interesting quasi-topological theory which has flavours of both Kodaira-Spencer and Donaldson-Thomas theory \cite{Garcia-Fernandez:2018ypt}. A reduction of this theory to three dimensions opens the door to more explicit computations, such as, for example, its partition function on the three-sphere. We leave this to future work. 

\section*{Acknowledgements}
We thank Rodrigo Barbosa, David Morrison, and Ethan Torres for interesting discussions and useful comments. The work of BSA and EES is supported by a grant from the Simons Foundation \#488569 (Bobby Acharya).
AK is supported by the Simons Foundation Grant \#488629 (Morrison) and the Simons Collaboration on Special Holonomy in Geometry,  Analysis, and Physics. EES would like to thank the International Centre for Theoretical Physics Trieste for their hospitality during the bulk of the project.

\appendix

\section{An $N=2$ Solution: The t'Hooft-Polyakov Monopole}
\label{app:T'Hooft-Polyakov geometry}
Instead of the complex structure of the SYZ fibration, we can alternatively endow $\mathbb{R}^3\times T^3$ with the following complex structure,
\begin{align}
z^1&= x^1+i\,x^2\nonumber\\
z^2&= x^3+i\,\phi^1\nonumber\\
z^3&= \phi^2+i\,\phi^3\:.
\end{align}
That is, the coordinate $z^3$ is the complex coordinate of an elliptic curve. This is more reminiscent of the local geometric structure required by heterotic duality with F-theory, and may thus be important for studying the local nature of this duality. The duality is again done fiber wise, with the elliptic curve now playing the role of $T^3$ in the Hull-Strominger system reduction. We also point out that the reduced geometry we study in this example now preserves $N=2$ supersymmetry rather than $N=1$, and so a continuous deformation to the reduced system of a stable complex flat connection is unlikely to exist. 

A reduction of the holomorphic Yang-Mills equations to $\mathbb{R}^3$, assuming a flat connection on the elliptic curve spanned by $\{\phi^2,\phi^3\}$, then reduces to the equation
\begin{equation}
\label{eq:SUSYtHP}
F^x=*_3\d_{A^x}\psi\:,
\end{equation}
where we have defined $A^y=\psi(x)\d \phi^1$. This is precisely the equation satisfied by the t'Hooft-Polyakov monopole \cite{t1974magnetic, polyakov1996particle}. The reduction of the Bianchi identity \eqref{eq:BI} now becomes
\begin{equation}
\label{eq:PolEqDil}
*_3\Delta_{\d} e^\Phi=\tfrac{\a}{2}\tr\,(\d_{A^x}\psi\wedge*_3\d_{A^x}\psi)=\tfrac{\a}{2}\d*_3\left(\tr\,(\psi\d_{A^x}\psi)\right)=\tfrac{\a}{4}\d*_3\d\left(\tr\,\psi^2\right)
\end{equation}
where $\Phi$ is the dilaton, and the corresponding three-dimensional metric is conformally flat given by $g_{ij}=e^\Phi\delta_{ij}$. In the second equality, we have used \eqref{eq:SUSYtHP} and the Bianchi identity for $F^x$. Equation \eqref{eq:PolEqDil} is solved by
\begin{equation}
e^\Phi=-\tfrac{\a}{4}\,\tr\,\psi^2+C\:,
\end{equation}
for some constant $C$ which can be thought of as an overall volume modulus. 

The geometric system in this case is far simpler than the $\a$-corrected equations for a stable complex flat connection, and can indeed be solved exactly if we have solutions to \eqref{eq:SUSYtHP}, for example the exact solution of \cite{prasad1975exact}. This is perhaps not surprising due to the enhanced supersymmetry. 


\bibliographystyle{JHEP}

\begin{thebibliography}{10}

\bibitem{Candelas:1985en}
P.~Candelas, G.~T. Horowitz, A.~Strominger, and E.~Witten, {\it {Vacuum
  Configurations for Superstrings}},  {\em Nucl.Phys.} {\bf B258} (1985)
  46--74.

\bibitem{Strominger1986}
A.~{Strominger}, {\it {Superstrings with torsion}},  {\em Nuclear Physics B}
  {\bf 274} (Sept., 1986) 253--284.

\bibitem{Hull1986357}
C.~Hull, {\it Compactifications of the heterotic superstring},  {\em Physics
  Letters B} {\bf 178} (1986), no.~4 357 -- 364.

\bibitem{Acharya:2001gy}
B.~S. Acharya and E.~Witten, {\it {Chiral fermions from manifolds of G(2)
  holonomy}},  \href{http://arxiv.org/abs/hep-th/0109152}{{\tt
  hep-th/0109152}}.

\bibitem{Witten:1995ex}
E.~Witten, {\it {String theory dynamics in various dimensions}},  {\em Nucl.
  Phys. B} {\bf 443} (1995) 85--126,
  [\href{http://arxiv.org/abs/hep-th/9503124}{{\tt hep-th/9503124}}].

\bibitem{Strominger:1996it}
A.~Strominger, S.-T. Yau, and E.~Zaslow, {\it {Mirror symmetry is T duality}},
  {\em Nucl. Phys.} {\bf B479} (1996) 243--259,
  [\href{http://arxiv.org/abs/hep-th/9606040}{{\tt hep-th/9606040}}].

\bibitem{donaldson2017adiabatic}
S.~Donaldson, {\it Adiabatic limits of co-associative kovalev--lefschetz
  fibrations},  in {\em Algebra, geometry, and physics in the 21st century},
  pp.~1--29.
\newblock Springer, 2017.

\bibitem{Acharya:2002kv}
B.~S. Acharya, {\it {A Moduli fixing mechanism in M theory}},
  \href{http://arxiv.org/abs/hep-th/0212294}{{\tt hep-th/0212294}}.

\bibitem{Pantev:2009de}
T.~Pantev and M.~Wijnholt, {\it {Hitchin's Equations and M-Theory
  Phenomenology}},  {\em J. Geom. Phys.} {\bf 61} (2011) 1223--1247,
  [\href{http://arxiv.org/abs/0905.1968}{{\tt arXiv:0905.1968}}].

\bibitem{Braun:2018vhk}
A.~P. Braun, S.~Cizel, M.~Hübner, and S.~Schäfer-Nameki, {\it {Higgs bundles
  for M-theory on $G_{2}$-manifolds}},  {\em JHEP} {\bf 03} (2019) 199,
  [\href{http://arxiv.org/abs/1812.06072}{{\tt arXiv:1812.06072}}].

\bibitem{kovalev}
A.~Kovalev, {\it {Twisted connected sums and special Riemannian holonomy}},
  {\em J. Reine Angew. Math} {\bf 565} (2003) 125--160.

\bibitem{Corti:2012kd}
A.~Corti, M.~Haskins, J.~Nordstr\"om, and T.~Pacini, {\it
  {$\mathrm{G}_{2}$-manifolds and associative submanifolds via semi-Fano
  $3$-folds}},  {\em Duke Math. J.} {\bf 164} (2015), no.~10 1971--2092,
  [\href{http://arxiv.org/abs/1207.4470}{{\tt arXiv:1207.4470}}].

\bibitem{Corti2013}
A.~Corti, M.~Haskins, J.~Nordstr\"om, and T.~Pacini, {\it {Asymptotically
  cylindrical Calabi-Yau $3$-folds from weak Fano $3$-folds}},  {\em Geom.
  Topol.} {\bf 17} (2013), no.~4 1955--2059.

\bibitem{Hubner:2020yde}
M.~Hubner, {\it {Local $G_2$-Manifolds, Higgs Bundles and a Colored Quantum
  Mechanics}},  \href{http://arxiv.org/abs/2009.07136}{{\tt arXiv:2009.07136}}.

\bibitem{Barbosa:2019bgh}
R.~Barbosa, M.~Cveti\v{c}, J.~J. Heckman, C.~Lawrie, E.~Torres, and
  G.~Zoccarato, {\it {T-branes and $G_2$ backgrounds}},  {\em Phys. Rev. D}
  {\bf 101} (2020), no.~2 026015, [\href{http://arxiv.org/abs/1906.02212}{{\tt
  arXiv:1906.02212}}].

\bibitem{Cecotti:2010bp}
S.~Cecotti, C.~Cordova, J.~J. Heckman, and C.~Vafa, {\it {T-Branes and
  Monodromy}},  {\em JHEP} {\bf 07} (2011) 030,
  [\href{http://arxiv.org/abs/1010.5780}{{\tt arXiv:1010.5780}}].

\bibitem{Hitchin2001}
N.~Hitchin, {\it {Stable Forms and Special Metrics}},
  \href{http://arxiv.org/abs/math/0107101}{{\tt math/0107101}}.

\bibitem{Gauntlett:2003cy}
J.~P. Gauntlett, D.~Martelli, and D.~Waldram, {\it {Superstrings with intrinsic
  torsion}},  {\em Phys.Rev.} {\bf D69} (2004) 086002,
  [\href{http://arxiv.org/abs/hep-th/0302158}{{\tt hep-th/0302158}}].

\bibitem{Hull198651}
C.~Hull, {\it Anomalies, ambiguities and superstrings},  {\em Physics Letters
  B} {\bf 167} (1986), no.~1 51 -- 55.

\bibitem{Hull1986187}
C.~Hull and P.~Townsend, {\it World-sheet supersymmetry and anomaly
  cancellation in the heterotic string},  {\em Physics Letters B} {\bf 178}
  (1986), no.~2D3 187 -- 192.

\bibitem{Sen1986289}
A.~Sen, {\it (2, 0) supersymmetry and space-time supersymmetry in the heterotic
  string theory},  {\em Nuclear Physics B} {\bf 278} (1986), no.~2 289 -- 308.

\bibitem{Ivanov:2009rh}
S.~Ivanov, {\it {Heterotic supersymmetry, anomaly cancellation and equations of
  motion}},  {\em Phys.Lett.} {\bf B685} (2010) 190--196,
  [\href{http://arxiv.org/abs/0908.2927}{{\tt arXiv:0908.2927}}].

\bibitem{Martelli:2010jx}
D.~Martelli and J.~Sparks, {\it {Non-K\"{a}hler Heterotic Rotations}},  {\em
  Adv.Theor.Math.Phys.} {\bf 15} (2011) 131--174,
  [\href{http://arxiv.org/abs/1010.4031}{{\tt arXiv:1010.4031}}].

\bibitem{delaOssa:2014msa}
X.~de~la Ossa and E.~E. Svanes, {\it {Connections, Field Redefinitions and
  Heterotic Supergravity}},  \href{http://arxiv.org/abs/1409.3347}{{\tt
  arXiv:1409.3347}}.

\bibitem{Green1984117}
M.~B. Green and J.~H. Schwarz, {\it Anomaly cancellations in supersymmetric d =
  10 gauge theory and superstring theory},  {\em Physics Letters B} {\bf 149}
  (1984), no.~1D3 117 -- 122.

\bibitem{donaldson1985anti}
S.~K. Donaldson, {\it Anti self-dual yang-mills connections over complex
  algebraic surfaces and stable vector bundles},  {\em Proceedings of the
  London Mathematical Society} {\bf 50} (1985), no.~1 1--26.

\bibitem{uhlenbeck1986existence}
K.~Uhlenbeck and S.-T. Yau, {\it On the existence of hermitian-yang-mills
  connections in stable vector bundles},  {\em Communications on Pure and
  Applied Mathematics} {\bf 39} (1986), no.~S1 S257--S293.

\bibitem{t1974magnetic}
G.~t~Hooft, {\it Magnetic monopoles in unified theories},  {\em Nucl. Phys. B}
  {\bf 79} (1974), no.~CERN-TH-1876 276--284.

\bibitem{polyakov1996particle}
A.~M. Polyakov, {\it Particle spectrum in quantum field theory},  in {\em 30
  Years Of The Landau Institute?Selected Papers}, pp.~540--541.
\newblock World Scientific, 1996.

\bibitem{Acharya:1998pm}
B.~S. Acharya, {\it {M theory, Joyce orbifolds and superYang-Mills}},  {\em
  Adv. Theor. Math. Phys.} {\bf 3} (1999) 227--248,
  [\href{http://arxiv.org/abs/hep-th/9812205}{{\tt hep-th/9812205}}].

\bibitem{corlette1988flat}
K.~Corlette et~al., {\it Flat $ g $-bundles with canonical metrics},  {\em
  Journal of differential geometry} {\bf 28} (1988), no.~3 361--382.

\bibitem{gagliardo2012geometric}
M.~Gagliardo and K.~Uhlenbeck, {\it Geometric aspects of the kapustin--witten
  equations},  {\em Journal of Fixed Point Theory and Applications} (2012)
  1--14.

\bibitem{Barbosa:2019hts}
R.~Barbosa, {\it {A Deformation Family for Closed $G_2$-Structures on ADE
  Fibrations}},  \href{http://arxiv.org/abs/1910.10742}{{\tt
  arXiv:1910.10742}}.

\bibitem{Carlevaro:2008qf}
L.~Carlevaro, D.~Israel, and P.~Petropoulos, {\it {Double-Scaling Limit of
  Heterotic Bundles and Dynamical Deformation in CFT}},  {\em Nucl. Phys. B}
  {\bf 827} (2010) 503--544, [\href{http://arxiv.org/abs/0812.3391}{{\tt
  arXiv:0812.3391}}].

\bibitem{Carlevaro:2009jx}
L.~Carlevaro and D.~Israel, {\it {Heterotic Resolved Conifolds with Torsion,
  from Supergravity to CFT}},  {\em JHEP} {\bf 01} (2010) 083,
  [\href{http://arxiv.org/abs/0910.3190}{{\tt arXiv:0910.3190}}].

\bibitem{Halmagyi:2016pqu}
N.~Halmagyi, D.~Israel, and E.~E. Svanes, {\it {The Abelian Heterotic
  Conifold}},  {\em JHEP} {\bf 07} (2016) 029,
  [\href{http://arxiv.org/abs/1601.07561}{{\tt arXiv:1601.07561}}].

\bibitem{Halmagyi:2017lqm}
N.~Halmagyi, D.~Israel, M.~Sarkis, and E.~E. Svanes, {\it {Heterotic
  Hyper-K\"ahler flux backgrounds}},  {\em JHEP} {\bf 08} (2017) 138,
  [\href{http://arxiv.org/abs/1706.01725}{{\tt arXiv:1706.01725}}].

\bibitem{Anderson:2010mh}
L.~B. Anderson, J.~Gray, A.~Lukas, and B.~Ovrut, {\it {Stabilizing the Complex
  Structure in Heterotic Calabi-Yau Vacua}},  {\em JHEP} {\bf 1102} (2011) 088,
  [\href{http://arxiv.org/abs/1010.0255}{{\tt arXiv:1010.0255}}].

\bibitem{Anderson:2014xha}
L.~B. Anderson, J.~Gray, and E.~Sharpe, {\it {Algebroids, Heterotic Moduli
  Spaces and the Strominger System}},  {\em JHEP} {\bf 1407} (2014) 037,
  [\href{http://arxiv.org/abs/1402.1532}{{\tt arXiv:1402.1532}}].

\bibitem{delaOssa:2014cia}
X.~de~la Ossa and E.~E. Svanes, {\it {Holomorphic Bundles and the Moduli Space
  of N=1 Supersymmetric Heterotic Compactifications}},  {\em JHEP} {\bf 10}
  (2014) 123, [\href{http://arxiv.org/abs/1402.1725}{{\tt arXiv:1402.1725}}].

\bibitem{Garcia-Fernandez:2015hja}
M.~Garcia-Fernandez, R.~Rubio, and C.~Tipler, {\it {Infinitesimal moduli for
  the Strominger system and Killing spinors in generalized geometry}},
  \href{http://arxiv.org/abs/1503.07562}{{\tt arXiv:1503.07562}}.

\bibitem{delaOssa:2015maa}
X.~de~la Ossa, E.~Hardy, and E.~E. Svanes, {\it {The Heterotic Superpotential
  and Moduli}},  {\em JHEP} {\bf 01} (2016) 049,
  [\href{http://arxiv.org/abs/1509.08724}{{\tt arXiv:1509.08724}}].

\bibitem{Candelas:2016usb}
P.~Candelas, X.~de~la Ossa, and J.~McOrist, {\it {A Metric for Heterotic
  Moduli}},  \href{http://arxiv.org/abs/1605.05256}{{\tt arXiv:1605.05256}}.

\bibitem{McOrist:2016cfl}
J.~McOrist, {\it {On the Effective Field Theory of Heterotic Vacua}},
  \href{http://arxiv.org/abs/1606.05221}{{\tt arXiv:1606.05221}}.

\bibitem{Ashmore:2018ybe}
A.~Ashmore, X.~De~La~Ossa, R.~Minasian, C.~Strickland-Constable, and E.~E.
  Svanes, {\it {Finite deformations from a heterotic superpotential:
  holomorphic Chern-Simons and an $L_\infty$ algebra}},  {\em JHEP} {\bf 10}
  (2018) 179, [\href{http://arxiv.org/abs/1806.08367}{{\tt arXiv:1806.08367}}].

\bibitem{Garcia-Fernandez:2018emx}
M.~Garcia-Fernandez, R.~Rubio, C.~Shahbazi, and C.~Tipler, {\it {Canonical
  metrics on holomorphic Courant algebroids}},
  \href{http://arxiv.org/abs/1803.01873}{{\tt arXiv:1803.01873}}.

\bibitem{Garcia-Fernandez:2018ypt}
M.~Garcia-Fernandez, R.~Rubio, and C.~Tipler, {\it {Holomorphic string
  algebroids}},  {\em Trans. Am. Math. Soc.} {\bf 373} (2020), no.~10
  7347--7382, [\href{http://arxiv.org/abs/1807.10329}{{\tt arXiv:1807.10329}}].

\bibitem{Garcia-Fernandez:2020awc}
M.~Garcia-Fernandez, R.~Rubio, and C.~Tipler, {\it {Gauge theory for string
  algebroids}},  \href{http://arxiv.org/abs/2004.11399}{{\tt
  arXiv:2004.11399}}.

\bibitem{prasad1975exact}
M.~Prasad and C.~M. Sommerfield, {\it Exact classical solution for the't hooft
  monopole and the julia-zee dyon},  {\em Physical Review Letters} {\bf 35}
  (1975), no.~12 760.

\end{thebibliography}

\providecommand{\href}[2]{#2}\begingroup\raggedright\endgroup

\end{document}